\documentclass[lettersize,journal]{IEEEtran}

\usepackage{cite}
\usepackage{amsmath}
\usepackage{amsmath,amssymb,amsfonts}
\usepackage{algorithm}
\usepackage{algpseudocode}
\usepackage{multirow}
\usepackage{graphicx}
\usepackage{textcomp}
\usepackage{xcolor}
\usepackage{colortbl}
\usepackage{array}
\usepackage{pdfpages}
\usepackage{textcomp}
\usepackage{enumerate}
\usepackage{xfrac}
\usepackage{amssymb}
\usepackage{mathtools}
\usepackage{amsmath}
\usepackage{enumerate}
\usepackage{blindtext}
\usepackage{enumitem}
\usepackage{graphicx}
\usepackage{caption}
\usepackage{subcaption}
\graphicspath{{Figures/}}
\usepackage[justification=centering]{caption}
\usepackage{setspace}
\usepackage{bbm}
\usepackage{bm}
\usepackage{optidef}
\usepackage {comment} 
\usepackage{listings}
\usepackage{float}
\usepackage{pgfplots}
\usepackage{pgfplotstable}
\pgfplotsset{compat=1.15}
\usepackage{xcolor}
\pgfplotsset{every tick label/.append style={font=\tiny}}

\usepackage{hyperref}
\hypersetup{
	colorlinks=true,
	linkcolor=blue,
	filecolor=blue,   
	citecolor=blue,  
	urlcolor=blue,
	pdftitle={Deep Reinforcement Learning-Aided Strategies for Big Data Offloading in Vehicular Networks},
	pdfpagemode=FullScreen,
}
\usepackage{tikz}
\usepackage{amsmath}
\usepackage{amsfonts}
\usepackage{dblfloatfix}  
\usetikzlibrary{positioning}
\usetikzlibrary{positioning, shapes.geometric}
\usetikzlibrary{positioning, arrows.meta}
\usetikzlibrary{decorations.pathreplacing}
\usetikzlibrary{calc}
\usepgfplotslibrary{fillbetween}

\definecolor{figcolor1}{RGB}{228,26,28} 
\definecolor{figcolor2}{RGB}{55,126,184}   
\definecolor{figcolor3}{RGB}{77,175,74}   
\definecolor{figcolor4}{RGB}{152,78,163}  
\definecolor{figcolor5}{RGB}{255,127,0}

\begin{document}
    \title{Deep Reinforcement Learning-Aided Strategies for Big Data Offloading in Vehicular Networks}
	\author{\IEEEauthorblockN{Talha Aky{\i}ld{\i}z and Hessam Mahdavifar\\} 	
		\thanks{T. Aky{\i}ld{\i}z is with the EECS Dept., University of Michigan, Ann Arbor, MI, 48104, USA (email: akyildiz@umich.edu).}
		\thanks{H. Mahdavifar is with the EECS Dept., University of Michigan, Ann Arbor, MI, 48104, USA  and ECE Dept., Northeastern University, Boston, MA, 02115, USA (email: h.mahdavifar@northeastern.edu).}
		\thanks{Part of this work was presented at the IEEE 97th Vehicular Technology Conference on Communications (VTC2023-Spring) \cite{akyildiz2023}.}
		\thanks{This research was supported in part by Ford Motor Company and National Science Foundation (NSF) under Grant CCF-2312752.}
		
	}
	
	\maketitle
	\normalsize
	
	\begin{abstract}	
	We consider vehicular networking scenarios where existing vehicle-to-vehicle (V2V) links can be leveraged for an effective uploading of large-size data to the network. In particular, we consider a group of vehicles where one vehicle can be designated as the \textit{leader} and other \textit{follower} vehicles can offload their data to the leader vehicle or directly upload it to the base station (or a combination of the two). In our proposed framework, the leader vehicle is responsible for receiving the data from other vehicles and processing it in order to remove the redundancy (deduplication) before uploading it to the base station. We present a mathematical framework of the considered network and formulate two separate optimization problems for minimizing (i) total time and (ii) total energy consumption by vehicles for uploading their data to the base station. We employ deep reinforcement learning (DRL) tools to obtain solutions in a dynamic vehicular network where network parameters (e.g., vehicle locations and channel coefficients) vary over time. Our results demonstrate that the application of DRL is highly beneficial, and data offloading with deduplication can significantly reduce the time and energy consumption. Furthermore, we present comprehensive numerical results to validate our findings and compare them with alternative approaches to show the benefits of the proposed DRL methods.
	\end{abstract}
	
	
	\vspace{-0.3cm}
	\section{Introduction}
	With the rapid advancement of mobile communication technologies, vehicles are now evolved into internet devices, called internet of vehicles (IoVs). These vehicles are equipped with various resources and components empowering them to provide services for their surroundings. These resources comprise sensors embedded within the vehicle, communication interfaces, and on-board units featuring computing and storage capabilities. Leveraging these features, IoVs function as mobile service providers, offering a wide range of services such as data storage, task computation and cloud services \cite{xu2018internet,ji2020survey}. Such capabilities contribute vehicles to produce massive IoV data which requires a suitable and well-designed communication between vehicles and also their surroundings, e.g., road side units (RSUs), base stations and pedestrians.
	
	Thanks to the advancements in vehicular communications with the emerging and developing 5G communication techniques, IoVs are able to exchange information with any entity, employing vehicle-to-everything (V2X) communications. This evolution brings a vast range of benefits for traffic and data management, e.g., autonomous driving, safety enhancement, traffic efficiency, and social infotainment \cite{garcia2021tutorial}. With the rapid expansion of these applications, vehicular ad-hoc networks (VANETs) have appeared as an ideal solution for vehicular communications by allowing vehicles to communicate with its surroundings directly by providing communication and processing capabilities to the vehicles. V2X can offer a wide range of communication types, e.g., vehicle-to-vehicle (V2V), and vehicle-to-infrastructure (V2I) \cite{zeadally2020vehicular}. 
	
	As vehicular data continues to proliferate with technological improvements, the implementation of an efficient offloading mechanism becomes imperative for vehicular data upload to the network for enhancing the quality of vehicle applications, leading to benefits such as reduced communication time and energy consumption \cite{zhou2018data}. Within VANETs, there exists two significant forms of offloading mechanisms, V2V and V2I. V2V offloading leverages collaborative efforts among neighboring vehicles through V2V communication, while, V2I offloading enhances uploading capabilities by transferring IoV data to the network infrastructure equipped with edge servers. In comparison, V2V offloading can offer less communication delay than that of V2I offloading especially for scenarios where the proximity between vehicles is generally closer compared to the distance between a vehicle and network infrastructure.
	
	In this paper, we consider scenarios where existing V2V links can be leveraged for an effective offloading of data to the network, i.e., the base station, especially data of types that are not time-sensitive, e.g., camera photos that will be used to update navigation systems. Moreover, vehicles in close proximity experience similar type of road conditions and traffic, and hence, the large-size data (images, videos) generated by the vehicles will also have similar contents. V2V communication can be really beneficial in such scenarios, enabling vehicles to execute deduplication prior to transmitting their data to the network infrastructure \cite{al2022promise}. On the other hand, without V2V links, each vehicle needs to upload its data to the base station or cloud with a lot of redundant information. This brings two undesirable outcomes: 1) A large cellular network traffic, and 2) Energy/time inefficiency over the vehicular network. However, we envision scenarios where vehicles can exchange data between each other through V2V links and remove/mitigate redundancy before uploading it to the base station, thereby avoiding congested network scenarios and providing energy/time efficiency.
	
	While existing literature has explored various aspects of vehicular data offloading, there remains a significant gap in addressing the optimization of time and energy consumption for big data offloading in VANETs with deduplication capabilities. To the best of our knowledge, no prior work has comprehensively addressed the challenge of dynamically distributing data between V2V and V2I links while considering redundancy elimination, leader vehicle selection, and the application of various deep reinforcement learning algorithms in time-varying vehicular environments. This paper presents a novel framework that bridges this gap by introducing a chunk-based data offloading approach with intelligent deduplication and adaptive resource allocation.
	
	The main contributions of this paper are summarized as
	\begin{itemize}	
		\item We present a comprehensive system model for vehicular data offloading that incorporates chunk-based data segmentation, enabling fine-grained offloading decisions. The model includes network establishment and architecture, communication models for both V2V and V2I links, a leader selection algorithm, and a deduplication framework that quantifies redundancy elimination benefits.
		
		\item We develop a mathematical framework and formulate two distinct optimization problems aimed at minimizing (i) total time and (ii) total energy consumption for data uploading in dynamic VANETs. Our formulation considers key decision variables including transmission power allocation and per-chunk offloading fractions, subject to constraints on latency, energy budgets, and power limits.
		
		\item We propose a deep reinforcement learning framework that addresses the inherent uncertainty and dynamics of vehicular environments. We implement and compare three different DRL algorithms in both centralized and decentralized configurations, providing insights into their relative performance for vehicular offloading.
		
		\item We present comprehensive numerical results demonstrating the effectiveness of our approach under various network conditions and redundancy levels. Our evaluations show that DRL-based offloading with deduplication can achieve significant reductions in both time and energy consumption compared to baseline strategies.
	\end{itemize}
	
	The remainder of this paper is organized as follows. Section II reviews related work on vehicle data offloading in VANETs. Section III presents our system model. Section IV formulates the optimization problems for time and energy minimization. Section V details our deep reinforcement learning approach with DRL algorithm implementations. Section VI provides simulation parameters and comprehensive numerical results. Finally, Section VII concludes the paper with key findings.	
	
	\section{Related Works}
	The authors in \cite{ancona2014performance} provide the performance boundaries for different aggregation scenarios by utilizing V2V offloading of floating car data (FCD), i.e., small-sized messages, to relieve cellular networks from FCD traffic. The problem formulation and optimization using graph theory for data forwarding/relaying through V2V and intermittent connectivity of access points are studied in \cite{kolios2014extract}. To improve the quality of service (QoS) provisioning, the work in \cite{saleem2021qos} proposes QoS-aware data offloading via established V2V and V2I links by providing traffic classification and overload control over the network. 
	
	In the context of networks facilitated by mobile edge computing (MEC), \cite{huang2018v2v} presents a routing algorithm to find a long-lasting path between vehicles through V2V for a software-defined network (SDN) to reduce cellular traffic by offloading. The dynamic nature of vehicular networks is addressed in \cite{huang2020k} and \cite{huang2020delay}, which propose k-hop-limited offloading approaches that adapt to changing network conditions. The influence of dynamic user behaviors and vehicle mobility patterns on offloading performance is investigated in \cite{du2022sdn} for edge and cloud computing systems. In \cite{liu2023mobility} and \cite{fan2023joint}, it is shown how mobility-aware multi-hop task offloading and joint V2I/V2V resource allocation strategies can effectively address the challenges of dynamic vehicular environments and varying network parameters in VANETs.
	
	Regarding big data-related V2V data offloading, the impacts and challenges provided by the big data for the VANETs are reviewed in detail in \cite{cheng2018big}. In \cite{zhou2017social}, a social big data content dissemination offloading approach through V2V links is proposed and a mathematical formulation of an optimization problem is also presented. A cooperative fog computing where local fog servers can share resources between each other with an aim of improving energy efficiency for big data applications is studied in \cite{zhang2017cooperative}. Another line of works in \cite{zhao2019computation,du2018computation,wang2019computation} have studied computation offloading in VANETs under different considerations. A comprehensive survey considering V2V, V2I, and V2X-based data offloading can be found in \cite{zhou2018data}.
	
	In the line of game theory studies in vehicular networks, two distinct WiFi offloading mechanisms based on game theory are presented to offload cellular traffic is proposed in \cite{cheng2016opportunistic}. It is demonstrated that both mechanisms contribute to enhancing the overall performance of WiFi offloading and achieving better fairness among vehicles. In \cite{su2016game}, the authors study the scenario where parked vehicles can serve as content provider under Stackelberg game. The work in \cite{aujla2017data} presents a Stackelberg game framework for data offloading for multiple followers within an SDN. Another related work in \cite{yang2019stackelberg} formulates a Stackelberg game  and derives a Nash equilibrium to find an optimal service provider. 
	
	The integration of artificial intelligence (AI) and machine learning (ML) techniques has emerged as a crucial enabler for next-generation vehicular networks. The comprehensive survey in \cite{du2020machine} examines machine learning techniques for wireless networks, exploring how to carry forward enhanced bandwidth, massive access, and ultrareliable service capabilities. Beyond traditional ML approaches, advanced techniques including federated learning for collaborative vehicular data processing \cite{posner2021federated}, graph neural networks for wireless communication applications \cite{lee2022graph}, and deep reinforcement learning for vehicular edge computing \cite{zhang2022adaptive} have shown promise in addressing the challenges of dynamic vehicular environments.

	\begin{table*}[t]
		\caption{Summary of Important Notations}
		\centering
		\footnotesize
		\renewcommand{\arraystretch}{1.0}
		\begin{tabular}{c|l|c|l}
			\hline
			\textbf{Notation} & \textbf{Description} & \textbf{Notation} & \textbf{Description} \\
			\hline
			$N(t)$ & Number of vehicles in the cluster at time $t$ & $B^{v2v}$ & Communication bandwidth of V2V links \\
			$K_i$ & Number of chunks for vehicle $i$ &  $B^{v2i}$ & Communication bandwidth of V2I links \\
			$\delta_{i,t}$ & Offloading fraction for chunk in slot $t$ & $R_{ij}^{v2v}[t]$ & V2V rate between vehicle $i$ and $j$ in slot $t$ \\
			$\beta_{i,t}$ & Redundancy ratio of chunk from vehicle $i$ in slot $t$ & $R_{i}^{v2i}[t]$ & V2I rate between vehicle $i$ and BS in slot $t$ \\
			$d_{i,k}$ & Size of chunk $k$ from vehicle $i$ & $p_i^{v2v}[t]$ & V2V transmission power of vehicle $i$ in slot $t$ \\
			$h_{ij}[t]$ & Channel gain between vehicle $i$ and $j$ in slot $t$ &  $p_i^{v2i}[t]$ & V2I transmission power of vehicle $i$ in slot $t$ \\			
			$g_{i}[t]$ & Channel gain between vehicle $i$ and BS in slot $t$ & $p_{\max}$ & Maximum transmission power \\
			\hline
		\end{tabular}
		\label{table:Notation}
        \vspace{-0.5cm}
	\end{table*}
	
	Among these AI/ML techniques, deep reinforcement learning (DRL) has gained particular significance in vehicular communications and task/data offloading as it empowers vehicles with the ability to make intelligent decisions in dynamic and complex environments. In \cite{ye2019deep}, decentralized resource allocation mechanisms are proposed, leveraging DRL to optimize sub-band and power level allocation for V2V transmissions. In \cite{liu2019deep}, a vehicle-assisted offloading scheme is introduced for computational tasks within a vehicular network, employing DRL to optimize computation offloading and resource allocation. In \cite{ning2019deep}, the authors present a three-layer offloading framework within IoV by addressing two sub-problems and solving them with the aid of DRL. The study in \cite{zhu2020multiagent} unveils a multi-agent DRL approach designed for vehicular computation offloading in MEC. The application of DRL extends to various domains, including channel state information compression in VANETs \cite{wang2020learn}, joint beam allocation and relay selection for millimeter-wave vehicular networks \cite{ju2023deep}, optimal beam association for vehicular networks \cite{van2021optimal}, and fast mmWave V2I beam alignment by leveraging vehicle position information and DRL \cite{qiao2024deep}. DRL has also been used in UAV-enabled mobile communications and UAV traffic management (UTM) systems \cite{rizvi2024multi,chintareddy2025federated}. These works demonstrate the versatility of DRL and learning-based methods in addressing diverse challenges in wireless communication systems, from physical layer optimization to high-level resource allocation. A comprehensive survey in \cite{mekrache2022deep} reveals recent advances and future trends of DRL techniques for vehicular networks.

	\vspace{-0.3cm}
	\section{System Model}
	In this section, we present a comprehensive system model for vehicular data offloading. We begin by describing the network establishment process, detailing how vehicles form clusters. We then introduce the network architecture, explaining the roles  of leader and follower vehicles. Next, we present the chunk-based data segmentation approach and detail the communication model for both V2V and V2I links. Subsequently, we describe the leader vehicle selection algorithm and present the deduplication model that captures how redundant data removed at the leader vehicle. A compact summary of the notation introduced is provided in Table~\ref{table:Notation}.
	
	\vspace{-0.3cm}
	\subsection{Network Establishment}
	In our proposed model, each vehicle within the VANET ecosystem is equipped with both 5G and WiFi technologies. The inclusion of 5G introduces two essential interfaces. Firstly, the PC5 interface, commonly referred to as sidelink, facilitates direct interaction between vehicles enabling V2V communication. Secondly, the Uu interface, operating as the downlink/uplink, allows vehicles to establish connections with the network infrastructure, the base station.
	
	With the aid of the PC5 interface, vehicles can establish V2V links with the peers in close proximity, forming a robust vehicular network. In addition to 5G technology, WiFi is utilized for efficient data transfer between vehicles via the established V2V links. This dual connectivity approach enables vehicles to leverage both 5G and WiFi capabilities for enhanced communication and data exchange within the VANET environment. Once V2V links have been established between vehicles, the formation of a vehicle cluster becomes feasible. We note that the scope of a cluster is constrained by the extent of WiFi coverage, ensuring that all vehicles within the cluster can maintain effective communication. It is noteworthy that clusters may undergo periodic updates as a result of dynamic conditions, which may include factors such as the arrival or departure of vehicles from the cluster.

	\vspace{-0.4cm}
	\subsection{Network Architecture}
	\label{subsec:network-architecture}
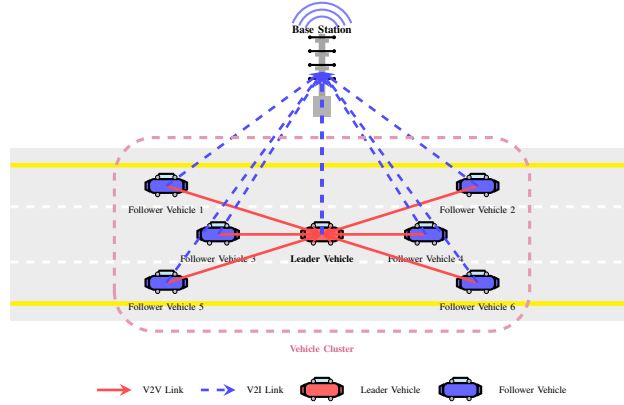
\begin{figure*}[!t]
		\centering
		\begin{tikzpicture}[scale=0.46, transform shape]
			\tikzstyle{v2v}=[->, >=stealth, thick, red!70, line width=1.0pt]
			\tikzstyle{v2i}=[->, >=stealth, thick, blue!70, dashed, line width=1.0pt]
			\tikzstyle{cluster}=[draw, dashed, rounded corners=15pt, purple!40, line width=1.2pt]
			
			\def\car#1#2#3#4{
				\begin{scope}[shift={(#1,#2)}, rotate=#3]
					\filldraw[draw=black, fill=#4, thick, rounded corners=2pt] (-0.5,-0.2) rectangle (0.5,0.2);
					\filldraw[draw=black, fill=#4, thick] (0.5,-0.15) -- (0.6,-0.1) -- (0.6,0.1) -- (0.5,0.15) -- cycle;
					\filldraw[draw=black, fill=#4, thick] (-0.5,-0.15) -- (-0.6,-0.1) -- (-0.6,0.1) -- (-0.5,0.15) -- cycle;
					\filldraw[draw=black, fill=cyan!20, thick] (-0.3,0.2) -- (-0.2,0.35) -- (0.2,0.35) -- (0.3,0.2) -- cycle;
					\filldraw[black] (-0.25,-0.25) circle (0.06);
					\filldraw[black] (0.25,-0.25) circle (0.06);
					\filldraw[black] (-0.25,0.25) circle (0.06);
					\filldraw[black] (0.25,0.25) circle (0.06);
					\filldraw[gray!50] (-0.25,-0.25) circle (0.03);
					\filldraw[gray!50] (0.25,-0.25) circle (0.03);
					\filldraw[gray!50] (-0.25,0.25) circle (0.03);
					\filldraw[gray!50] (0.25,0.25) circle (0.03);
				\end{scope}
			}
			
			\fill[gray!15] (-9,-2.5) rectangle (9,2.5);
			\draw[white, dashed, line width=0.04cm] (-9,-0.8) -- (9,-0.8);
			\draw[white, dashed, line width=0.04cm] (-9,0.8) -- (9,0.8);
			\draw[yellow, line width=0.06cm] (-9,2.0) -- (9,2.0);
			\draw[yellow, line width=0.06cm] (-9,-2.0) -- (9,-2.0);
			
			\begin{scope}[shift={(0,4.0)}]
				\fill[gray!60] (-0.25,-0.6) rectangle (0.25,0);
				\draw[line width=2.5pt, gray!70] (0,0) -- (0,1.8);
				\draw[line width=1.5pt, gray!70] (-0.2,0.4) -- (0.2,0.4);
				\draw[line width=1.5pt, gray!70] (-0.2,0.8) -- (0.2,0.8);
				\draw[line width=1.5pt, gray!70] (-0.2,1.2) -- (0.2,1.2);
				\draw[line width=1.5pt, gray!70] (-0.2,1.6) -- (0.2,1.6);
				\foreach \y in {0.5,0.9,1.3,1.7} {
					\draw[thick, black] (-0.35,\y) -- (0.35,\y);
					\filldraw[black] (-0.35,\y) circle (0.04);
					\filldraw[black] (0.35,\y) circle (0.04);
				}
				\foreach \r in {0.5,0.7,0.9} {
					\draw[blue!50, thick] (0,1.8) ++(20:\r) arc (20:160:\r);
				}
				\node[above, font=\small\bfseries] at (0,1.7) {Base Station};
			\end{scope}
			\coordinate (bs) at (0,4.7);
			
			\car{0}{0}{0}{red!60}
			\coordinate (leader) at (0,0);
			\node[font=\footnotesize, below=2pt] at (0,-0.4) {\textbf{Leader Vehicle}};
			
			\car{-4.5}{1.4}{0}{blue!60}
			\coordinate (f1) at (-4.5,1.4);
			\node[font=\footnotesize, below=2pt] at (-4.5,1.0) {Follower Vehicle 1};
			
			\car{4.5}{1.4}{0}{blue!60}
			\coordinate (f2) at (4.5,1.4);
			\node[font=\footnotesize, below=2pt] at (4.5,1.0) {Follower Vehicle 2};
			
			\car{-3.0}{0}{0}{blue!60}
			\coordinate (f3) at (-3.0,0);
			\node[font=\footnotesize, below=2pt] at (-3.0,-0.4) {Follower Vehicle 3};
			
			\car{3.0}{0}{0}{blue!60}
			\coordinate (f4) at (3.0,0);
			\node[font=\footnotesize, below=2pt] at (3.0,-0.4) {Follower Vehicle 4};
			
			\car{-4.5}{-1.4}{0}{blue!60}
			\coordinate (f5) at (-4.5,-1.4);
			\node[font=\footnotesize, below=2pt] at (-4.5,-1.8) {Follower Vehicle 5};
			
			\car{4.5}{-1.4}{0}{blue!60}
			\coordinate (f6) at (4.5,-1.4);
			\node[font=\footnotesize, below=2pt] at (4.5,-1.8) {Follower Vehicle 6};
			
			\draw[v2v] (f1) -- (leader);
			\draw[v2v] (f2) -- (leader);
			\draw[v2v] (f3) -- (leader);
			\draw[v2v] (f4) -- (leader);
			\draw[v2v] (f5) -- (leader);
			\draw[v2v] (f6) -- (leader);
			
			\draw[v2i] (f1) -- (bs);
			\draw[v2i] (f2) -- (bs);
			\draw[v2i] (f3) -- (bs);
			\draw[v2i] (f4) -- (bs);
			\draw[v2i] (f5) -- (bs);
			\draw[v2i] (f6) -- (bs);
			
			\draw[v2i, line width=1.5pt] (leader) -- (bs);
			
			\draw[cluster] (-6.0,-2.8) rectangle (6.0,2.8);
			\node[font=\footnotesize\bfseries, purple!60] at (0,-3.3) {Vehicle Cluster};
			
			\begin{scope}[shift={(0,-4.5)}]
				\draw[v2v] (-6.5,0) -- (-5.5,0);
				\node[right, font=\footnotesize] at (-5.3,0) {V2V Link};
				
				\draw[v2i] (-3.5,0) -- (-2.5,0);
				\node[right, font=\footnotesize] at (-2.3,0) {V2I Link};
				
				\car{0}{0}{0}{red!60}
				\node[right, font=\footnotesize] at (1.0,0) {Leader Vehicle};
				
				\car{4}{0}{0}{blue!60}
				\node[right, font=\footnotesize] at (5,0) {Follower Vehicle};
			\end{scope}
			
		\end{tikzpicture}
		\caption{An illustrative scenario of vehicles organized into a single cluster on a three-lane road. One vehicle is designated as the leader, receiving chunks from follower vehicles over V2V links, performing deduplication, and ultimately uploading unique data to the base station.}
		\label{figusecase}
        \vspace{-0.5cm}
	\end{figure*}

	We consider a group of vehicles within an urban area, navigating roadways at low to medium speeds. We assume these vehicles share their data either with nearby peers or with the base station, and we organize vehicles in close proximity into clusters for efficient and reliable communication. We assume that within each cluster, only one vehicle can be appointed as the \emph{leader}, while the remaining vehicles function as \emph{followers}. Fig. \ref{figusecase} illustrates a sample roadway scenario, in which a selected leader vehicle and multiple follower vehicles can communicate through V2V and V2I links.

    In our model, the leader vehicle is responsible for receiving data from the follower vehicles within the cluster, performing \emph{deduplication} to remove redundant content, and then uploading the resulting unique data to the base station. Conversely, each follower vehicle can either (i)~upload its data directly to the base station or (ii)~offload it to the leader vehicle for possible deduplication (or use a combination of both). The former approach may cause congestion in the cellular network if vehicles transfer large volumes of data individually, leading to increased time and energy usage. By contrast, offloading data to the leader can leverage the existing V2V links between vehicles in close proximity, thereby reducing duplication and alleviating the cellular network load. 

	\vspace{-0.2cm}
	\subsection{Chunk-Based Data Segmentation}
    \label{subsec:chunk-segmentation}
    In our vehicular network, each vehicle $i$ holds large amounts of data $D_i$, and transmitting the entire data object as a single monolithic block can be inefficient under different channel conditions. To introduce finer control over how data is handled, we adopt a chunk-based approach in which the total data $D_i$ is partitioned into a sequence of $K_i$ smaller chunks, i.e., 
    \begin{equation}
    \label{eq:chunk-partition}
    D_i = \sum_{k=1}^{K_i} d_{i,k},
    \end{equation}
    where $d_{i,k}$ is the size of the $k$-th chunk from vehicle $i$. 
    
    We adopt a discrete-time framework where time is divided into slots $t \in \{1, 2, ..., T\}$, each with duration $\Delta t$. This time-slot structure enables us to model how chunks are transmitted over time-varying vehicular channels, where each chunk transmission occurs within a single slot. We assume that chunk sizes are chosen such that each chunk can be completely transmitted within a single time slot.

	Each chunk $d_{i,k}$ can be further subdivided into data payload and additional overhead, reflecting the metadata and control information needed to manage individual chunks. Denoting the payload portion by $d_{i,k}^{\text{data}}$ and the per-chunk overhead portion by $d_{i,k}^{\text{meta}}$, we have $d_{i,k} = d_{i,k}^{\text{data}} \oplus d_{i,k}^{\text{meta}}$, where $\oplus$ denotes concatenation. The overhead typically encompasses identifiers (chunk indices and source information), size parameters, and hash signatures that allow for integrity verification or deduplication checks. We note that $d_{i,k}^{\text{meta}}$ is usually small compared to $d_{i,k}^{\text{data}}$, but it is essential for managing the chunked data effectively.

    In our discrete-time model chunk-based offloading process, each follower vehicle $i$ divides its data into $K_i$ chunks. Then, for each chunk $k$, it selects an offloading fraction $\delta_{i,k} \in [0,1]$. This fraction determines what portion of chunk $d_{i,k}$ is sent to the leader via V2V, while the remaining portion $1-\delta_{i,k}$ is uploaded directly to the base station via V2I. This allows vehicles to optimize their offloading strategy based on channel conditions. For notational simplicity, we assume continuous sequential transmission where vehicle $i$ transmits chunk $k$ in time slot $t = k$. This enables us to use $\delta_{i,t}$ and $d_{i,t}$ as shorthand for $\delta_{i,k}$ and $d_{i,k}$ respectively, where $k = t$. 
	
	\vspace{-0.26cm}
	\subsection{Communication Model}
    We now present the communication model in the vehicular environment for the established V2V and V2I links. Our approach incorporates a stochastic channel model to account for signal propagation between vehicles and the base station. Notably, this model does not necessitate precise information regarding the geometric arrangement of vehicles, and it permits channel parameters to be stochastically determined via extensive real-life measurements. The characteristics of the propagation model encompass various components, including path loss, signal attenuation concerning distance, large-scale propagation, signal fading due to environmental obstacles, and small-scale propagation, impact of multipath propagation.
	
	We select log-normal shadowing model to define channel characteristics\footnote{We note that it is possible to consider different channel modeling approaches, e.g., geometry-based (ray-tracing). However, these models generally require detailed information about the environment, hence, we adopt stochastic model as it is more suitable for non-specific scenarios \cite{boban2014geometry}.} for V2V link and assume flat fading within each time slot. Let $f_{ij}[t]$ be the small-scale propagation coefficient between vehicles $i$ and $j$ in time slot $t$. Then, we define $h_{ij}[t]$, the channel gain between vehicles $i$ and $j$ in time slot $t$, which is given as 
	\begin{equation}
		h_{ij}[t] =  \left|f_{ij}[t]\right|^2 \nu_0^{v2v} \left(\frac{d_0}{d_{ij}[t]}\right)^{\gamma^{v2v}}, 
	\end{equation}
	where $d_0$ is the reference distance, and $d_{ij}[t]$ is the distance between vehicles $i$ and $j$ in time slot $t$. $\nu_0^{v2v}$ is the path loss constant and $\gamma^{v2v}$ is the path loss exponent of V2V links. 

	We consider orthogonal resource allocation where each V2V link is assigned dedicated frequency or time resources, eliminating inter-vehicle interference. This modeling approach aligns with modern vehicular communication standards, e.g., 3GPP C-V2X and IEEE 802.11p \cite{naik2019ieee}, which employ scheduling mechanisms and resource allocation schemes to ensure interference-free communication. Under this consideration, the transmission rate of V2V links between vehicle $i$ and $j$ in time slot $t$ is given by
	\begin{equation}
		R_{ij}^{v2v}[t] = B^{v2v} \log_2 \left(1+\frac{p_i^{v2v}[t] h_{ij}[t]}{N_0 B^{v2v}}\right),
	\end{equation}
    where $B^{v2v}$ is the communication bandwidth of V2V links, $p_i^{v2v}[t]$ is the transmission power of vehicle $i$ over V2V links and $N_0$ is the noise power spectral density.
	
	Similarly, we derive the channel gain and transmission rate between vehicles and the base station. We denote the channel gain between vehicle $i$ and the base station by $g_{i}[t]$ in time slot $t$, which can be given as 
	\begin{equation}
		g_{i}[t] =  |f_{i}[t]|^2 \nu_0^{v2i} \left(\frac{d_0}{d_{i}[t]}\right)^{\gamma^{v2i}}, 
	\end{equation}
	where $d_i[t]$ and $f_i[t]$ are the distance and the small-scale propagation coefficient between follower vehicle $i$ and the base station in slot $t$, respectively. Similar to the V2V case, we assume orthogonal resource allocation for V2I links. The transmission rate between vehicle $i$ and the base station, i.e., 
	\begin{equation}
		R_{i}^{v2i}[t] = B^{v2i} \log_2 \left(1+\frac{p_i^{v2i}[t]  g_{i}[t]}{N_0 B^{v2i}}\right),
	\end{equation}
    where $B^{v2i}$ is the communication bandwidth and $p_i^{v2i}[t]$ represents the transmit power of vehicle $i$ for V2I links.

	\subsection{Leader Vehicle Selection}
	\label{subsec:leader-selection}
	Following the formation of each cluster, we select a single leader vehicle $i^*_t$ at time slot $t$ to coordinate offloading and perform deduplication on the data received from follower vehicles. In particular, the selected leader should simultaneously (i) provide strong V2V links to follower vehicles, (ii) maintain a favorable V2I rate to the base station for uploading unique chunks, and (iii) possess sufficient processing power to handle deduplication tasks without becoming a bottleneck. The leader vehicle selection criterion for time slot $t$ is given by
	\begin{equation}
		\label{eq:leader-dynamic}
		i^*_t = \underset{i \in \mathcal{N}}{\mathrm{argmax}} \Biggl[ \sum_{\substack{j=1, \\ j\neq i}}^{N(t)} 
		R_{ij}^{\mathrm{v2v}}[t] + R_{i}^{\mathrm{v2i}}[t] + \zeta  f_i \Biggr],
	\end{equation}
	where the first term represents the total V2V communication rate from all other vehicles, the second term is the V2I  rate from vehicle $i$ to the base station, and $f_i$ is the CPU frequency of vehicle $i$. The constant $\zeta \ge 0$ is a tunable weighting factor. 

	While dynamic leader selection could enable adaptation to changing network conditions, frequent leader switching would disrupt ongoing deduplication processes and require significant coordination overhead. Therefore, we adopt a stable leader selection strategy where the leader is selected at the beginning of the transmission period when $t = 0$ and remains fixed throughout the data transmission, $i^*_t = i^*_0 = i^*$ for all $t$.
	
	\subsection{Deduplication Model}
	\label{subsec:dedup-model}	
	We consider an analytical model for content-defined sub-chunking based deduplication at the leader vehicle to identify and eliminate redundant data within each time slot. Our approach extends conventional hash-based techniques  enabling partial redundancy detection at sub-chunk granularity.

	Each offloaded data chunk $\mathcal{D}_{j,t}$ of size $\delta_{j,t} d_{j,t}$ bytes from vehicle $j$ in slot $t$ is partitioned into sub-chunks using content-defined chunking (CDC). Let $\mathcal{S}_{j,t} = \{s_{j,t}^{(1)}, s_{j,t}^{(2)}, \ldots, s_{j,t}^{(m_{j,t})}\}$ denote the set of sub-chunks, where $m_{j,t} = |\mathcal{S}_{j,t}|$ is the number of sub-chunks produced. Let $\bar{L}$ denote the average sub-chunk length in slot $t$, i.e., $\bar{L} \approx \delta_{j,t} d_{j,t}/m_{j,t}$. The sub-chunks form a complete partition of the offloaded chunk, satisfying
	\begin{equation}
	\mathcal{D}_{j,t} = s_{j,t}^{(1)} \oplus s_{j,t}^{(2)} \oplus \cdots \oplus s_{j,t}^{(m_{j,t})},  \quad |\mathcal{D}_{j,t}| = \sum_{i=1}^{m_{j,t}} |s_{j,t}^{(i)}|,
	\end{equation}
	where $\oplus$ denotes concatenation operation. Each sub-chunk $s_{j,t}^{(i)}$ is processed through a cryptographic hash function (e.g., SHA-256) to generate a unique fingerprint
	\begin{equation}
	h_{j,t}^{(i)} = \text{Hash}(s_{j,t}^{(i)}), \quad \text{where } h_{j,t}^{(i)} \in \{0,1\}^{256}.
	\end{equation}
	The leader vehicle maintains a hash table $\mathcal{H}_t$ containing fingerprints of all unique sub-chunks received in slot $t$. For each incoming chunk, the redundancy ratio is computed as
	\begin{equation}
	\beta_{j,t} = \frac{\sum_{i=1}^{m_{j,t}} \mathbb{I}_{h_{j,t}^{(i)} \in \mathcal{H}_t} \cdot |s_{j,t}^{(i)}|}{\sum_{i=1}^{m_{j,t}} |s_{j,t}^{(i)}|},
	\end{equation}
	where $\mathbb{I}_{h_{j,t}^{(i)} \in \mathcal{H}_t}$ is the indicator function that equals 1 if sub-chunk $i$ is duplicate (i.e., its hash exists in $\mathcal{H}_t$) and 0 otherwise. This length-weighted formulation ensures $\beta_{j,t} \in [0,1]$ represents the fraction of duplicate bytes within the chunk from vehicle $j$ in slot $t$.

	In our discrete-time model, the leader processes incoming chunks sequentially. Let $D^r[t]$ denote the total data received by the leader via V2V links in slot $t$, and $D^u[t]$ denote the unique data remaining after deduplication that must be uploaded to the base station, i.e., 
 	\begin{equation}
 	D^r[t] = \sum_{\substack{j=1, \\j\neq i^*}}^{N(t)} \delta_{j,t} \cdot d_{j,t}, \quad
 	D^u[t] = \sum_{\substack{j=1, \\j\neq i^*}}^{N(t)} [1 - \beta_{j,t}] \cdot \delta_{j,t} \cdot d_{j,t},
 	\end{equation}
 	where $\delta_{j,t}$ and $d_{j,t}$ represent the offloading fraction and size of the chunk transmitted by vehicle $j$ in slot 
	$t$, and $\beta_{j,t}$ is the redundancy ratio computed via content-defined sub-chunking.

	The deduplication process comprises three primary operations with the following computational complexities \cite{cormen2009introduction}: (i) fingerprinting and cryptographic hashing of sub-chunks, with complexity $O(D^r[t])$, (ii) hash table lookup operations for sub-chunks with complexity $O(D^r[t]/\bar{L})$, and (iii) chunk management overhead for metadata processing with complexity $O(N_c[t])$. Using the computation model \cite{guo2019energy}, we obtain the time consumption for deduplication as
 	\begin{equation}
	T_{i^*}^{c}[t] = \frac{1}{f_{i^*}}(\underbrace{C_1 D^r[t]}_{\text{CDC \& hashing}} + \underbrace{C_2 \frac{D^r[t]}{\bar{L}}}_{\text{hash lookup}} + \underbrace{C_3 N_c[t]}_{\text{metadata}}),
 	\end{equation}
	where $C_1$ represents CPU cycles per bit for chunking and hash computation, $C_2$ captures CPU cycles per sub-chunk for hash table probe, $C_3$ represents per-chunk overhead for metadata management, $N_c[t] = N(t) - 1$ is the number of chunks, and $f_{i^*}$ is the CPU frequency of the leader vehicle.

	\noindent For compactness, we can obtain the simplified form
 	\begin{equation}
	T_{i^*}^{c}[t] = \frac{1}{f_{i^*}} \left(C_4 D^r[t] + C_3 (N(t) - 1)\right),
 	\end{equation}
	where $C_4 = C_1 + C_2/\bar{L}$ represents the effective CPU cycles per bit for all operations, i.e., CDC, hashing, and hash table lookups. The energy consumption following the time consumption is given by
 	\begin{equation}
	E_{i^*}^{c}[t] = \kappa f_{i^*}^2 T_{i^*}^{c}[t] + P_{\text{static}} T_{i^*}^{c}[t] \approx \kappa f_{i^*}^2 T_{i^*}^{c}[t],
 	\end{equation}
 	where $\kappa$ is the dynamic power coefficient, and $P_{\text{static}}$ represents static power consumption. For processors 
	operating at maximum frequency, the dynamic component dominates, allowing us to use the simplified form.

	\section{Problem Formulation}
	\label{sec:prob-formulation}
	We now formulate the optimization problem for vehicular data offloading. The key challenge is to determine the optimal offloading strategy that minimizes either the total time or energy consumption while ensuring all data chunks are successfully transmitted. This involves making intelligent decisions about how to split each chunk between V2V and V2I links, taking into account the dynamic channel conditions, deduplication benefits, and computational overhead.
	
	In our discrete-time model, each vehicle transmits one chunk per time slot. For vehicle $i$ in time slot $t$, recall that $\delta_{i,t}$ is the fraction of chunk $d_{i,t}$ sent to the leader vehicle via V2V, and $(1-\delta_{i,t})$ is sent directly to the base station via V2I. We denote the V2V rate between follower vehicle $i$ and leader vehicle $i^*$ by $R_{i i^*}^{\mathrm{v2v}}[t]$ and the V2I rate between vehicle $i$ and the base station by $R_{i}^{\mathrm{v2i}}[t]$, respectively. The time to transmit the V2V and V2I portions of the chunk in slot $t$ is
	\begin{equation}
		T_{i}^{\mathrm{v2v}}[t] = \frac{\delta_{i,t} \cdot d_{i,t}}{R_{i i^*}^{\mathrm{v2v}}[t]}, \qquad
		T_{i}^{\mathrm{v2i}}[t] = \frac{(1-\delta_{i,t}) \cdot d_{i,t}}{R_{i}^{\mathrm{v2i}}[t]}.
	\end{equation}

	Since V2V and V2I transmissions can occur in parallel, the time for vehicle $i$ to complete its chunk transmission in slot $t$ is determined by the slower of the two paths, i.e.,
	\begin{equation}
		T_{i}^{\mathrm{trans}}[t] = \max\{T_{i}^{\mathrm{v2v}}[t],\,T_{i}^{\mathrm{v2i}}[t]\}.
	\end{equation}

	In each slot $t$, the leader vehicle $i^*$ performs deduplication on the received chunks. The deduplication processing time is $T_{i^*}^{c}[t]$. The time required for the leader to upload the unique data $D^u[t]$ from slot $t$ to the base station is
	\begin{equation}
		T_{i^*}^{\mathrm{v2i}}[t] = \frac{D^u[t]}{R_{i^*}^{\mathrm{v2i}}[t]}.
	\end{equation}

	From an energy perspective, the energy consumed by vehicle $i$ in slot $t$ for V2V and V2I transmissions is computed by multiplying the transmission power by the corresponding transmission time. This gives us
	\begin{equation}
		E_{i}^{\mathrm{v2v}}[t] = p_i^{\mathrm{v2v}}[t] \cdot T_{i}^{\mathrm{v2v}}[t], \quad
		E_{i}^{\mathrm{v2i}}[t] = p_i^{\mathrm{v2i}}[t] \cdot T_{i}^{\mathrm{v2i}}[t].
	\end{equation}
	Since both V2V and V2I links operate simultaneously and consume power independently, the total transmission energy for vehicle $i$ in slot $t$ is
	\begin{equation}
		E_i^{\mathrm{trans}}[t] = E_{i}^{\mathrm{v2v}}[t] + E_{i}^{\mathrm{v2i}}[t].
	\end{equation}

	At the leader vehicle $i^*$, the energy consumed for deduplication in slot $t$ is $E_{i^*}^{c}[t]$. The energy required for the leader to upload the unique data $D^u[t]$ to the base station is
	\begin{equation}
		E_{i^*}^{\mathrm{v2i}}[t] = p_{i^*}^{\mathrm{v2i}}[t] \cdot T_{i^*}^{\mathrm{v2i}}[t].
	\end{equation}

	We now present the objective functions that are solved independently for each time slot $t$. The objective functions capture the total system cost in terms of time and energy, accounting for all operations in the cluster. Specifically, let
	\begin{align}
		F_{\mathrm{time}}[t] &= \sum_{\substack{i=1,\, \\ i\neq i^*}}^{N(t)} T_i^{\mathrm{trans}}[t] + T_{i^*}^{c}[t] + T_{i^*}^{\mathrm{v2i}}[t], \\
		F_{\mathrm{energy}}[t] &= \sum_{\substack{i=1,\, \\ i\neq i^*}}^{N(t)} E_i^{\mathrm{trans}}[t] + E_{i^*}^{c}[t] + E_{i^*}^{\mathrm{v2i}}[t],
	\end{align}
	where the first terms represent the cumulative transmission cost from all follower vehicles, while the second and third terms account for the leader's deduplication processing and unique data upload costs to the base station, respectively.

	The optimization problem aims to determine the optimal offloading fractions and power allocations for all follower vehicles in each slot, enabling vehicles to optimize their strategies to minimize either total completion time or total energy consumption. For each time slot $t$, we solve the following, 
	\begin{mini!}|s|[2] 
		{\substack{%
		\delta_{i,t},
		p_i^{\mathrm{v2v}}[t], p_i^{\mathrm{v2i}}[t]}}
		{F_{\mathrm{time}}[t] \text{ or } F_{\mathrm{energy}}[t]}
		{\label{eq:unified-problem}}
		{}
		\addConstraint{\delta_{i,t} \in [0,1]}{, \forall i \neq i^*}
		\addConstraint{T_i^{\mathrm{trans}}[t] \le T_{\max}}{, \forall i\neq i^*}
		\addConstraint{T_{i^*}^{\mathrm{v2i}}[t]+T_{i^*}^{c}[t] \le T_{\max}}{}
		\addConstraint{E_i^{\mathrm{trans}}[t] \le E_{\max}}{, \forall i\neq i^*}
		\addConstraint{E_{i^*}^{\mathrm{v2i}}[t]+E_{i^*}^{c}[t] \le E_{\max}}{}
		\addConstraint{p_i^{\mathrm{v2v}}[t] \le p_{\max}}{, \forall i \neq i^*}
		\addConstraint{p_i^{\mathrm{v2i}}[t] \le p_{\max}}{, \forall i \neq i^*},
	\end{mini!}
	where the first constraint restricts the offloading fraction to the interval $[0,1]$ for all vehicles. The second and third constraints enforce time budgets for follower vehicles and the leader within each slot. The fourth and fifth constraints enforce energy budgets for follower vehicles and the leader. The final two constraints limit V2V and V2I transmit powers separately, allowing each link to utilize the full power budget.

	The optimization problem exhibits several properties that make traditional solution methods inadequate. The objective functions are non-convex due to the $\max$ operator in transmission time calculations and fractional terms involving data rates. The decision variables are coupled through the deduplication process, preventing decomposition into independent subproblems.  While convex optimization tools cannot handle the non-convexity, and game-theoretic equilibria are intractable to compute in real-time for this problem, deep reinforcement learning offers a viable alternative. DRL can navigate non-convex optimization landscapes, learn adaptive policies through interaction, and scale to high-dimensional action spaces without requiring analytical solutions.

	\vspace{-0.2cm}
	\section{Deep Reinforcement Learning}
	\label{sec:RL}
	In this section, we present a deep reinforcement learning framework to solve the vehicular offloading optimization problem. Given the computational intractability of traditional methods for this coupled and time-varying problem, we leverage ability of DRL to learn effective policies without requiring analytical solutions or convexity assumptions.

	We begin by defining the key components of our DRL framework: state space design that captures network conditions, action space including offloading decisions and power control, and reward functions that encode our optimization objectives with constraint violations. We consider three distinct DRL algorithms: deep Q-network (DQN), deep deterministic policy gradient (DDPG), and soft actor-critic (SAC). Each algorithm is implemented in both centralized and decentralized configurations. Finally, we provide comprehensive training procedures and implementation details of our framework.
	
	\vspace{-0.5cm}
	\subsection{DRL Framework for Vehicular Offloading}
	We formulate the vehicular offloading problem as a sequential decision-making process suitable for deep reinforcement learning, where agents learn optimal strategies through continuous interaction with the dynamic environment. During each time slot $t$, agents observe the vehicular network state, make decisions about offloading fractions and power allocations, and receive reward signals that align with our optimization objectives either minimizing time or energy.

	The key components of our DRL framework are given as
	\begin{itemize}
		\item \textbf{State space} $\mathcal{S}$: The set of observations available to agents, including channel conditions, vehicle positions, and system status.
		\item \textbf{Action space} $\mathcal{A}$: The set of control decisions, comprising offloading fractions $\delta_{i,t}$ and power allocations $p_i^{v2v}[t]$ and $p_i^{v2i}[t]$ for each vehicle.
		\item \textbf{Reward function} $\mathcal{R}$: The feedback signal that encodes our optimization objectives  and constraint violations.
	\end{itemize}

	Through repeated interactions, agents learn policies $\pi: \mathcal{S} \rightarrow \mathcal{A}$ that map observed states to actions, aiming to minimize the expected cumulative reward over time. We implement this framework in both centralized and decentralized frameworks.

	\subsubsection{State Space Design}
	We design distinct state representations for centralized and decentralized frameworks, enabling comparison between global coordination and distributed decision-making approaches.

	\paragraph{Centralized State Space}
	In the centralized configuration, a single agent maintains global observability over the entire vehicular cluster. This approach provides complete visibility into all follower channel conditions, vehicle positions, and time slot information, enabling direct coordination through joint decision-making. The comprehensive state representation allows the agent to exploit system-wide patterns and dependencies for optimal offloading strategies. The centralized state vector $\mathbf{s}_t^{\text{c}} \in \mathbb{R}^{3N(t)-1}$ is  defined as
	\begin{equation}
	\begin{aligned}
	\mathbf{s}_t^{\text{c}} = [ &\log_{10}(g_1[t]), \ldots, \log_{10}(g_{N(t)-1}[t]), \\
	&\log_{10}(h_{1i^*}[t]), \ldots, \log_{10}(h_{(N(t)-1)i^*}[t]), \\
	&d_1[t]/d_{\text{norm}}, \ldots, d_{N(t)}[t]/d_{\text{norm}}, t/T ]^T,
	\end{aligned}
	\end{equation}
	where the components represent: (i) logarithmically scaled V2I channel gains for all follower vehicles to ensure numerical stability during training, (ii) logarithmically scaled V2V channel gains between each follower and the leader vehicle $i^*$, (iii) normalized distances from the base station for all $N(t)$ vehicles (including the leader), where $d_{\text{norm}}$ is a distance normalization constant, and (iv) normalized time slot progression within the episode of duration $T$.
	
	\paragraph{Decentralized State Space}
	In the decentralized configuration, each follower vehicle operates as an independent agent with limited local observability. Agents can only observe their own channel conditions, positions, and chunk information. This distributed architecture promotes scalability, and reduces communication overhead. Each follower vehicle $i$ maintains a compact local state vector $\mathbf{s}_{i,t}^{\text{d}} \in \mathbb{R}^6$ defined as
	\begin{equation}
	\begin{aligned}
	\mathbf{s}_{i,t}^{\text{d}} = [ &\log_{10}(g_i[t]), \log_{10}(h_{ii^*}[t]), d_i[t]/d_{\text{norm}}, d_{i^*}[t]/d_{\text{norm}}, \\
	&t/T, r_{\text{sys}}^{t-1} ]^T,
	\end{aligned}
	\end{equation}
	where the components include: (i) logarithmically scaled V2I channel gain between vehicle $i$ and the base station, (ii) logarithmically scaled V2V channel gain between vehicle $i$ and the leader $i^*$, (iii-iv) normalized distances of vehicle $i$ and the leader from the base station, (v) normalized time slot within the episode, and (vi) normalized previous system reward $r_{\text{sys}}^{t-1}$ providing implicit coordination feedback.

	\subsubsection{Action Space Design}
	The action space directly maps to the optimization variables namely, the offloading fraction $\delta_{i,t} \in [0,1]$ that determines the portion of each chunk sent via V2V, and the transmission powers $p_i^{v2v}[t]$ and $p_i^{v2i}[t]$ for V2V and V2I links respectively, enabling DRL agents to control data distribution and power allocation decisions.

	\paragraph{Centralized Action Space}
	In the centralized framework, a single agent controls all follower vehicles through joint action selection. The centralized agent leverages complete system visibility to make coordinated decisions, accounting for how actions of each vehicle affect overall time or energy consumption. The centralized agent outputs a high-dimensional joint action vector $\mathbf{a}_t^{\text{c}} \in \mathbb{R}^{3(N(t)-1)}$ defined as 
	\begin{equation}
	\small
	\mathbf{a}_t^{\text{c}} = \Big[ \delta_{1,t}, p_1^{\text{v2v}}[t], p_1^{\text{v2i}}[t], \ldots, \delta_{N(t)-1,t}, p_{N(t)-1}^{\text{v2v}}[t], p_{N(t)-1}^{\text{v2i}}[t] \Big]^T.
	\end{equation}

	\paragraph{Decentralized Action Space}
	In the decentralized approach, each follower vehicle functions as an autonomous decision-maker, selecting actions based solely on local observations. This distributed architecture promotes scalability and robustness by enabling parallel decision-making without requiring communication overhead or centralized coordination. Each follower vehicle $i$ independently selects action vector $\mathbf{a}_{i,t}^{\text{d}} \in \mathbb{R}^3$, i.e.,
	\begin{equation}
	\mathbf{a}_{i,t}^{\text{d}} = \begin{bmatrix} \delta_{i,t}, & p_i^{\text{v2v}}[t], & p_i^{\text{v2i}}[t] \end{bmatrix}^T.
	\end{equation}

	\subsubsection{Reward Function Design}
	The reward function translates the optimization objectives into learning signals for the DRL agents. Since reinforcement learning seeks to maximize rewards, we define the reward as the negative of the objective function with constraint penalties.

	We first define the base reward as the negative of the objective functions $F_{\text{time}}[t]$ and $F_{\text{energy}}[t]$. This transformation converts our minimization problems into maximization problems suitable for reinforcement learning, where agents seek to maximize cumulative rewards. The base rewards are can be given as
	\begin{align}
	\small
	r_t^{\text{time,base}} &= -\sum_{\substack{i=1, \\ i \neq i^*}}^{N(t)} T_i^{\text{trans}}[t] - T_{i^*}^{c}[t] - T_{i^*}^{\text{v2i}}[t], \\
	r_t^{\text{energy,base}} &=  -\sum_{\substack{i=1, \\ i \neq i^*}}^{N(t)} E_i^{\text{trans}}[t] - E_{i^*}^{c}[t] - E_{i^*}^{\text{v2i}}[t].
	\end{align}

	The complete reward functions incorporate constraint violations through penalty terms to ensure feasible solutions. The full reward structure becomes
	\begin{equation}
	\resizebox{\columnwidth}{!}{$
	\begin{aligned}
	r_t^{\text{time}} &= r_t^{\text{time,base}} - \lambda_{\text{cons}} \sum_{\substack{i=1, \\ i \neq i^*}}^{N(t)} \left(\mathbb{I}_i^{\text{FT}}[t] + \mathbb{I}_i^{\text{FE}}[t]\right) - \lambda_{\text{cons}} \left(\mathbb{I}^{\text{LT}}[t] + \mathbb{I}^{\text{LE}}[t]\right), \\
	r_t^{\text{energy}} &= r_t^{\text{energy,base}} - \lambda_{\text{cons}} \sum_{\substack{i=1, \\ i \neq i^*}}^{N(t)} \left(\mathbb{I}_i^{\text{FT}}[t] + \mathbb{I}_i^{\text{FE}}[t]\right) - \lambda_{\text{cons}} \left(\mathbb{I}^{\text{LT}}[t] + \mathbb{I}^{\text{LE}}[t]\right),
	\end{aligned}
	$}
	\end{equation} 
	where $\lambda_{\text{cons}} > 0$ is the constraint penalty coefficient. The indicator functions equal 1 when the corresponding constraint is violated and 0 otherwise. Specifically, $\mathbb{I}_i^{\text{FT}}[t]$ indicates whether the transmission time of follower $i$ exceeds $T_{\max}$, $\mathbb{I}^{\text{LT}}[t]$ indicates whether the combined processing and upload time of the leader exceeds $T_{\max}$, $\mathbb{I}_i^{\text{FE}}[t]$ indicates whether the transmission energy of follower $i$ exceeds $E_{\max}$, and $\mathbb{I}^{\text{LE}}[t]$ indicates whether the combined processing and upload energy of the leader exceeds $E_{\max}$.
	
	\vspace{-0.3cm}
	\subsection{Deep Reinforcement Learning Algorithms}
	We present the deep reinforcement learning algorithms employed to solve the vehicular offloading optimization problem. We first implement DQN to establish a baseline using value-based learning. Despite requiring action space discretization, sample efficiency and stable convergence of DQN make it well-suited for initial policy development and performance benchmarking. For continuous control, we employ DDPG, an actor-critic method that directly optimizes continuous actions without discretization. DDPG extends Q-learning to continuous domains through deterministic policy gradients, enabling precise control of power levels and offloading fractions. Finally, we implement SAC, which incorporates entropy regularization into the actor-critic framework. This approach encourages stochastic policies that explore the action space more effectively while mitigating value overestimation.

	\subsubsection{Deep Q-Network (DQN)}
	DQN represents a value-based reinforcement learning approach that approximates the state-action value function using deep neural networks. DQN operates on discrete action space, and it provides limited granularity compared with continuous control approaches.

	DQN learns a Q-function $Q_\theta(\mathbf{s}_t,\mathbf{a}_t)$ that estimates the expected cumulative reward for taking action $\mathbf{a}_t$ in state $\mathbf{s}_t$, where $\theta$ represents the neural network parameters. Here, $\mathbf{s}_t$ represents either the global state $\mathbf{s}_t^{\text{c}}$ in the centralized or the local state $\mathbf{s}_{i,t}^{\text{d}}$ in the decentralized configuration. Similarly, $\mathbf{a}_t$ denotes either the joint action vector $\mathbf{a}_t^{\text{c}}$ for all followers in centralized or the individual action $\mathbf{a}_{i,t}^{\text{d}}$ in decentralized setting. The optimal policy greedily selects actions, i.e.,
	\begin{equation}
	\pi^*(\mathbf{s}_t) = \arg\max_{\mathbf{a}_t} Q_\theta(\mathbf{s}_t,\mathbf{a}_t).
	\end{equation}

	The Q-network is trained by minimizing the temporal difference error, given as 
	\begin{equation}
	L(\theta) = \mathbb{E} \left[ \left( y_t - Q_\theta(\mathbf{s}_t,\mathbf{a}_t) \right)^2 \right],
	\end{equation}
	where $y_t = r_t + \gamma \max_{\mathbf{a}_{t+1}} Q_{\theta^-}(\mathbf{s}_{t+1}, \mathbf{a}_{t+1})$ is the target value computed using a slowly updated target network with parameters $\theta^-$, and $\gamma \in [0,1]$ is the discount factor that determines the importance of future rewards.

	\subsubsection{Deep Deterministic Policy Gradient (DDPG)}
	DDPG extends the discrete framework to continuous action spaces by combining policy gradient methods with Q-learning. It is particularly effective for our problem as it requires continuous control for offloading fractions and power allocations.

	DDPG uses an actor-critic architecture where the actor $\mu_\phi(\mathbf{s}_t)$ generates deterministic actions and the critic $Q_\theta(\mathbf{s}_t,\mathbf{a}_t)$ evaluates state-action pairs. The critic is trained using
	\begin{equation}
	L_Q(\theta) = \mathbb{E} \left[ \left( y_t - Q_\theta(\mathbf{s}_t,\mathbf{a}_t) \right)^2 \right],
	\end{equation}
	where $y_t = r_t + \gamma Q_{\theta^-}(\mathbf{s}_{t+1}, \mu_{\phi^-}(\mathbf{s}_{t+1}))$. Here, $\theta^-$ and $\phi^-$ denote the parameters of the target critic and target actor networks, respectively, which are slowly updated copies of the main networks to ensure training stability. The actor is updated to maximize expected Q-values through the policy gradient.

	\subsubsection{Soft Actor-Critic (SAC)}
	SAC is an algorithm that maximizes both expected rewards and policy entropy, incorporating exploration directly into the optimization objective.

	The SAC objective function combines reward maximization with entropy regularization
	\begin{equation}
	J(\phi) = \mathbb{E} \left[ \sum_t \gamma^t (r_t + \alpha H(\pi_\phi(\cdot|\mathbf{s}_t))) \right],
	\end{equation}
	where $H(\pi_\phi(\cdot|\mathbf{s}_t)) = -\mathbb{E}_{\mathbf{a}_t \sim \pi_\phi}[\log \pi_\phi(\mathbf{a}_t|\mathbf{s}_t)]$ is the policy entropy, $\phi$ denotes the policy network parameters, and $\alpha > 0$ is the temperature parameter that balances reward and entropy.

	SAC uses dual Q-networks $Q_{\theta_1}(\mathbf{s}_t,\mathbf{a}_t)$ and $Q_{\theta_2}(\mathbf{s}_t,\mathbf{a}_t)$ to mitigate overestimation bias, where $\theta_1$ and $\theta_2$ are the parameters of the two critic networks. The policy is updated by minimizing
	\begin{equation}
	L_\pi(\phi) = \mathbb{E} \left[ \alpha \log \pi_\phi(\mathbf{a}_t|\mathbf{s}_t) - \min(Q_{\theta_1}, Q_{\theta_2})(\mathbf{s}_t,\mathbf{a}_t) \right].
	\end{equation}
	The entropy-regularized objective encourages exploration while the dual Q-networks address value overestimation.

	\subsubsection{Action Space Discretization}
	DQN requires discrete action spaces, which we address through a systematic preset-based discretization of the continuous action space. This approach enables DQN to explore meaningful combinations of offloading fractions and power allocations while avoiding the combinatorial explosion of naive grid-based discretization. Table \ref{tab:action_presets} presents the 25 action presets, where each preset defines $[\delta, p_{\text{v2v}}/p_{\max}, p_{\text{v2i}}/p_{\max}]$ with normalized power values.
	
	\begin{table}[t]
		\caption{DQN Action Presets for Vehicular Offloading}
		\centering
		\renewcommand{\arraystretch}{1.1}
		\resizebox{\columnwidth}{!}{%
		\begin{tabular}{l|c|c|c||c|c|c}
			\hline
			\textbf{Scenario} & \textbf{$\boldsymbol{\delta}$} & \textbf{$\boldsymbol{p_{\text{v2v}}/p_{\max}}$} & \textbf{$\boldsymbol{p_{\text{v2i}}/p_{\max}}$} & \textbf{$\boldsymbol{\delta}$} & \textbf{$\boldsymbol{p_{\text{v2v}}/p_{\max}}$} & \textbf{$\boldsymbol{p_{\text{v2i}}/p_{\max}}$} \\
			\hline
			\hline
			\textbf{Extreme} & 0.0 & 0.0 & 1.0 & 1.0 & 1.0 & 0.0 \\
			\hline
			\multirow{3}{*}{\textbf{V2I-Favorable}} & 0.1 & 1.0 & 1.0 & 0.1 & 0.6 & 0.8 \\
			& 0.2 & 1.0 & 1.0 & 0.2 & 0.6 & 0.8 \\
			& 0.3 & 1.0 & 1.0 & 0.3 & 0.6 & 0.8 \\
			\hline
			\multirow{3}{*}{\textbf{Balanced}} & 0.4 & 1.0 & 1.0 & 0.4 & 0.7 & 0.7 \\
			& 0.5 & 1.0 & 1.0 & 0.5 & 0.7 & 0.7 \\
			& 0.6 & 1.0 & 1.0 & 0.6 & 0.7 & 0.7 \\
			\hline
			\multirow{3}{*}{\textbf{Energy-Critical}} & 0.3 & 0.4 & 0.6 & 0.4 & 0.5 & 0.6 \\
			& 0.5 & 0.5 & 0.5 & 0.6 & 0.6 & 0.5 \\
			& 0.7 & 0.6 & 0.4 & & & \\
			\hline
			\multirow{3}{*}{\textbf{V2V-Favorable}} & 0.7 & 1.0 & 1.0 & 0.7 & 0.8 & 0.6 \\
			& 0.8 & 1.0 & 1.0 & 0.8 & 0.8 & 0.6 \\
			& 0.9 & 1.0 & 1.0 & 0.9 & 0.8 & 0.6 \\
			\hline
		\end{tabular}%
		}
		\label{tab:action_presets}
	\end{table}
	
	The preset design captures representative vehicular scenarios: extreme channel conditions, scenarios favoring V2V cooperation when vehicles are distant from infrastructure, scenarios favoring V2I transmission when vehicles are near the base station, balanced channel conditions, and energy-constrained scenarios requiring power management. This systematic coverage enables DQN to learn effective policies without exhaustive exploration of the continuous action space.

	\begin{algorithm}[t]
	\footnotesize
	\caption{DRL Training for Vehicular Offloading}\label{alg:drl-training}
	\begin{algorithmic}[1]
	\State \textbf{Initialize} DRL agent(s) with random network parameters
	\State \textbf{Initialize} experience replay buffer(s)
	\State \textbf{Initialize} target networks and exploration parameters
	\For{episode $= 1$ to $M$}
	    \State Reset environment: vehicle positions, velocities, channel gains
	    \State Generate chunk assignments with redundancy ratios $\beta_{i,k}$
	    \State Select leader vehicle $i^*$ using \eqref{eq:leader-dynamic}
	    \State Observe initial state(s): $\mathbf{s}_0^{\text{c}}$ (centralized) or $\{\mathbf{s}_{i,0}^{\text{d}}\}$ (decentralized)
	    \For{time slot $t = 0$ to $T-1$}
	        \If{centralized approach}
	            \State Select joint action $\mathbf{a}_t^{\text{c}}$ using current policy
	        \Else
	            \State Each follower selects individual action $\mathbf{a}_{i,t}^{\text{d}}$ 
	        \EndIf
	        \State Execute actions in environment:
	        \State \quad - Compute V2V rates $R_{ii^*}^{\text{v2v}}[t]$ and V2I rates $R_i^{\text{v2i}}[t]$
	        \State \quad - Transmit chunk fractions $\delta_{i,t}$ via V2V and $(1-\delta_{i,t})$ via V2I  
	        \State \quad - Perform deduplication at leader: compute $D^u[t]$ 
	        \State \quad - Update vehicle positions and channel gains
	        \State Compute objective function cost $F_{\text{time}}[t]$ or $F_{\text{energy}}[t]$
	        \State Observe next state(s) $\mathbf{s}_{t+1}^{\text{c}}$ or $\{\mathbf{s}_{i,t+1}^{\text{d}}\}$ and reward(s) $r_t^{\text{time}}$ or $r_t^{\text{energy}}$
	        \State Store transition(s) in replay buffer
	        \If{sufficient data in replay buffer}
	            \State Sample mini-batch and update network parameters
	            \State Update target networks (DQN every 100 time slots, DDPG/SAC soft updates)
	        \EndIf
	        \If{all chunks transmitted or time/energy budget exceeded}
	            \State Break
	        \EndIf
	    \EndFor
	    \State Update exploration parameters
	\EndFor
	\end{algorithmic}
	\end{algorithm}
	
	\subsection{Training Procedure and Implementation Details}
	The complete training framework integrates the DRL algorithms with our vehicular offloading environment, incorporating both centralized and decentralized learning settings. Algorithm~\ref{alg:drl-training} presents the training procedure where agents learn optimal offloading policies through episodic interactions with the vehicular environment, updating their networks based on observed rewards and environmental feedback.
	
	For all DRL algorithms, we employ neural networks with two hidden layers of 256 neurons each, using ReLU activation and layer normalization. The networks are optimized using the Adam optimizer with algorithm-specific learning rates: DQN uses $\alpha = 10^{-4}$, DDPG employs actor and critic learning rates $\alpha_{\text{actor}} = 10^{-5}$ and $\alpha_{\text{critic}} = 5 \times 10^{-4}$, while SAC uses actor, critic, and temperature learning rates $\alpha_{\text{actor}} = \alpha_{\text{critic}} = \alpha_{\text{temp}} = 10^{-4}$. Training begins after collecting 10,000 initial experiences to ensure stable learning. Network updates occur every timestep with a batch size of 256 samples. Target networks are updated every 100 steps for DQN, with soft updates using $\tau = 0.001$ for DDPG and $\tau = 0.005$ for SAC.

	DQN uses an $\epsilon$-greedy exploration strategy with $\epsilon$ starting at 0.3, decaying to 0.05 with decay rate 0.999. DDPG employs Ornstein-Uhlenbeck noise for continuous action exploration with parameters $\theta = 0.1$, $\sigma = 0.05$, and noise scale 0.1. SAC uses entropy-regularized exploration with initial temperature $\alpha = 0.05$ and automatic entropy tuning. All algorithms use a replay buffer size of 500,000 experiences and discount factor $\gamma = 0.99$. Gradient clipping is applied with maximum norm 1.0 for DQN and SAC, and 0.5 for DDPG. 
	
	\section{Numerical Results}
	\label{sec:results}
	We now present comprehensive numerical results to evaluate the performance of our proposed DRL-based vehicular offloading framework. We begin by describing the simulation environment and key parameters used in our experiments. Subsequently, we analyze the training performance of proposed DRL algorithms. Finally, we evaluate the trained models under various network conditions, examining their performance in terms of both time and energy optimization.

	\begin{table}[t]
		\caption{Simulation Parameters}
		\centering
		\footnotesize
		\renewcommand{\arraystretch}{1.1}
		\begin{tabular}{l|c}
			\hline
			\textbf{Parameter} & \textbf{Value} \\
			\hline
			\hline
			\multicolumn{2}{c}{\textbf{Network Configuration}} \\
			\hline
			Number of vehicles $N(t)$ & 5 (default), 3-7 \\
			Number of time slots $T$ & 30 \\
			Time slot duration $\Delta t$ & 1 s \\
			\hline
			\multicolumn{2}{c}{\textbf{Vehicular Environment}} \\
			\hline
			Road length & 1000 m \\
			Number of lanes & 3 \\

			Vehicle speed range & $\mathcal{U}[10, 15]$ m/s, (36--54 km/h) \\
			Base station position & [200, 0] m \\
			Vehicle initial positions & $\mathcal{U}[0, 50]$ m \\
			\hline
			\multicolumn{2}{c}{\textbf{Communication Parameters}} \\
			\hline
			V2V bandwidth $B^{v2v}$ & 10 MHz \\
			V2I bandwidth $B^{v2i}$ & 20 MHz \\
			Noise power density $N_0$ & $4 \times 10^{-21}$ W/Hz \\
			Maximum power $p_{\max}$ & 200 mW (23 dBm) \\
			\hline
			\multicolumn{2}{c}{\textbf{Channel Model Parameters}} \\
			\hline
			Reference distance $d_0$ & 1 m \\
			V2V path loss constant $\nu_0^{v2v}$ & $2 \times 10^{-5}$ \\
			V2I path loss constant $\nu_0^{v2i}$ & $2 \times 10^{-5}$ \\
			V2V path loss exponent $\gamma^{v2v}$ & 3.5 \\
			V2I path loss exponent $\gamma^{v2i}$ & 3.5 \\
			\hline
			\multicolumn{2}{c}{\textbf{Data and Redundancy Parameters}} \\
			\hline
			Chunks per vehicle $K_i$ & 30 \\
			Chunk size $d_{i,k}$ & 20 Mb \\
			Redundancy ratio $\beta_{i,k}$ & 0.5 (default), 0.3--0.7 \\
			\hline
			\multicolumn{2}{c}{\textbf{Deduplication and Computation}} \\
			\hline
			CPU cycles per bit $C_1$ & 10 cycles/bit \\
			Per-chunk overhead $C_4$ & $10^6$ cycles \\
			CPU frequency $f_{i}$ & 2.8 GHz \\
			Hardware constant $\kappa$ & $10^{-27}$ \\
			\hline
			\multicolumn{2}{c}{\textbf{Optimization Constraints}} \\
			\hline
			Time budget per slot $T_{\max}$ & 1 s \\
			Energy budget per slot $E_{\max}$ & 1 J \\
			\hline
		\end{tabular}
		\label{tab:simulation_params}
	\end{table}

	\begin{figure*}[!h]
		\centering
		\begin{subfigure}[b]{0.48\textwidth}
			\centering
			\begin{tikzpicture}[scale=0.8]
				\begin{axis}[
					width=\linewidth,
					height=6cm,
					xlabel={Episode},
					ylabel={Time (s)},
					xmin=0, xmax=5000,
					ymin=25, ymax=50,
					ytick={25,30,35,40,45,50},
					grid=both,
					grid style={line width=.1pt, draw=gray!20},
					major grid style={line width=.2pt,draw=gray!50},
					legend pos=north east,
					legend style={font=\footnotesize, fill=white, fill opacity=1, draw=none},
					line width=1pt,
				]
				\addplot[color=figcolor1, line width=1.5pt, mark=*, smooth] coordinates {
					(499,42.33) (999,36.85) (1499,34.88) (1999,33.89) (2499,33.74) (2999,32.99) (3499,32.88) (3999,33.19) (4499,32.79) (4999,32.23)
				};
				\addplot[color=figcolor2, line width=1.5pt, mark=x, smooth] coordinates {
					(499,42.59) (999,33.60) (1499,31.71) (1999,31.17) (2499,30.94) (2999,30.75) (3499,30.38) (3999,30.06) (4499,30.17) (4999,29.86)
				};
				\addplot[color=figcolor3, line width=1.5pt, mark=+, smooth] coordinates {
					(499,49.51) (999,45.62) (1499,38.88) (1999,36.86) (2499,36.86) (2999,36.98) (3499,35.80) (3999,36.29) (4499,35.73) (4999,36.39)
				};
				\legend{C-DQN, C-DDPG, C-SAC}
				\end{axis}
			\end{tikzpicture}
			\vspace{-0.2cm}
			\caption{Centralized training (time optimization)}
		\end{subfigure}
		\hfill
		\begin{subfigure}[b]{0.48\textwidth}
			\centering
			\begin{tikzpicture}[scale=0.8]
				\begin{axis}[
					width=\linewidth,
					height=6cm,
					xlabel={Episode},
					ylabel={Time (s)},
					xmin=0, xmax=5000,
					ymin=25, ymax=50,
					ytick={25,30,35,40,45,50},
					grid=both,
					grid style={line width=.1pt, draw=gray!20},
					major grid style={line width=.2pt,draw=gray!50},
					legend pos=north east,
					legend style={font=\footnotesize, fill=white, fill opacity=1, draw=none},
					line width=1pt,
					smooth,
				]
				\addplot[color=figcolor1, line width=1.5pt, mark=*, smooth] coordinates {
					(499,38.27) (999,33.04) (1499,31.95) (1999,31.51) (2499,31.07) (2999,30.94) (3499,30.75) (3999,30.44) (4499,30.24) (4999,30.05)
				};
				\addplot[color=figcolor2, line width=1.5pt, mark=x, smooth] coordinates {
					(499,40.12) (999,35.27) (1499,32.68) (1999,30.56) (2499,29.72) (2999,29.55) (3499,29.50) (3999,29.39) (4499,29.13) (4999,29.06)
				};
				\addplot[color=figcolor3, line width=1.5pt, mark=+, smooth] coordinates {
					(499,46.38) (999,36.62) (1499,32.64) (1999,31.38) (2499,30.77) (2999,30.60) (3499,30.25) (3999,30.09) (4499,30.21) (4999,29.92)
				};
				\legend{D-DQN, D-DDPG, D-SAC}
				\end{axis}
			\end{tikzpicture}
			\vspace{-0.2cm}
			\caption{Decentralized training (time optimization)}
		\end{subfigure}
		\\
		\begin{subfigure}[b]{0.48\textwidth}
			\centering
			\begin{tikzpicture}[scale=0.8]
				\begin{axis}[
					width=\linewidth,
					height=6cm,
					xlabel={Episode},
					ylabel={Energy (J)},
					xmin=0, xmax=5000,
					ymin=6, ymax=12,
					grid=both,
					grid style={line width=.1pt, draw=gray!20},
					major grid style={line width=.2pt,draw=gray!50},
					legend pos=north east,
					legend style={font=\footnotesize, fill=white, fill opacity=1, draw=none},
					line width=1pt,
					smooth,
				]
				\addplot[color=figcolor1, line width=1.5pt, mark=*, smooth] coordinates {
					(499,11.72) (999,10.37) (1499,9.36) (1999,9.19) (2499,9.26) (2999,9.06) (3499,9.15) (3999,9.14) (4499,9.05) (4999,8.94)
				};
				\addplot[color=figcolor2, line width=1.5pt, mark=x, smooth] coordinates {
					(499,11.17) (999,9.79) (1499,8.99) (1999,8.69) (2499,8.48) (2999,8.32) (3499,8.04) (3999,7.76) (4499,7.51) (4999,7.48)
				};
				\addplot[color=figcolor3, line width=1.5pt, mark=+, smooth] coordinates {
					(499,10.97) (999,9.88) (1499,9.48) (1999,9.31) (2499,9.27) (2999,9.10) (3499,8.96) (3999,8.94) (4499,8.80) (4999,8.86)
				};
				\legend{C-DQN, C-DDPG, C-SAC}
				\end{axis}
			\end{tikzpicture}
			\vspace{-0.2cm}
			\caption{Centralized training (energy optimization)}
		\end{subfigure}
		\hfill
		\begin{subfigure}[b]{0.48\textwidth}
			\centering
			\begin{tikzpicture}[scale=0.8]
				\begin{axis}[
					width=\linewidth,
					height=6cm,
					xlabel={Episode},
					ylabel={Energy (J)},
					xmin=0, xmax=5000,
					ymin=6, ymax=12,
					grid=both,
					grid style={line width=.1pt, draw=gray!20},
					major grid style={line width=.2pt,draw=gray!50},
					legend pos=north east,
					legend style={font=\footnotesize, fill=white, fill opacity=1, draw=none},
					line width=1pt,
					smooth,
				]
				\addplot[color=figcolor1, line width=1.5pt, mark=*, smooth] coordinates {
					(499,10.91) (999,9.50) (1499,8.97) (1999,8.86) (2499,8.80) (2999,8.79) (3499,8.74) (3999,8.70) (4499,8.67) (4999,8.65)
				};
				\addplot[color=figcolor2, line width=1.5pt, mark=x, smooth] coordinates {
					(499,10.85) (999,9.38) (1499,9.00) (1999,8.56) (2499,8.24) (2999,7.97) (3499,7.65) (3999,7.28) (4499,6.90) (4999,6.88)
				};
				\addplot[color=figcolor3, line width=1.5pt, mark=+, smooth] coordinates {
					(499,9.86) (999,8.91) (1499,8.02) (1999,7.34) (2499,6.86) (2999,6.69) (3499,6.44) (3999,6.37) (4499,6.41) (4999,6.30)
				};
				\legend{D-DQN, D-DDPG, D-SAC}
				\end{axis}
			\end{tikzpicture}
			\vspace{-0.2cm}
			\caption{Decentralized training (energy optimization)}
		\end{subfigure}
		\caption{Training performance of centralized (C-DQN, C-DDPG, C-SAC) and decentralized (D-DQN, D-DDPG, D-SAC) DRL algorithms for time and energy optimization objectives across 5000 episodes. Each curve shows the moving average of the respective cost metric during training.}
		\label{fig:training_curves}
		\vspace{-0.4cm}
	\end{figure*}
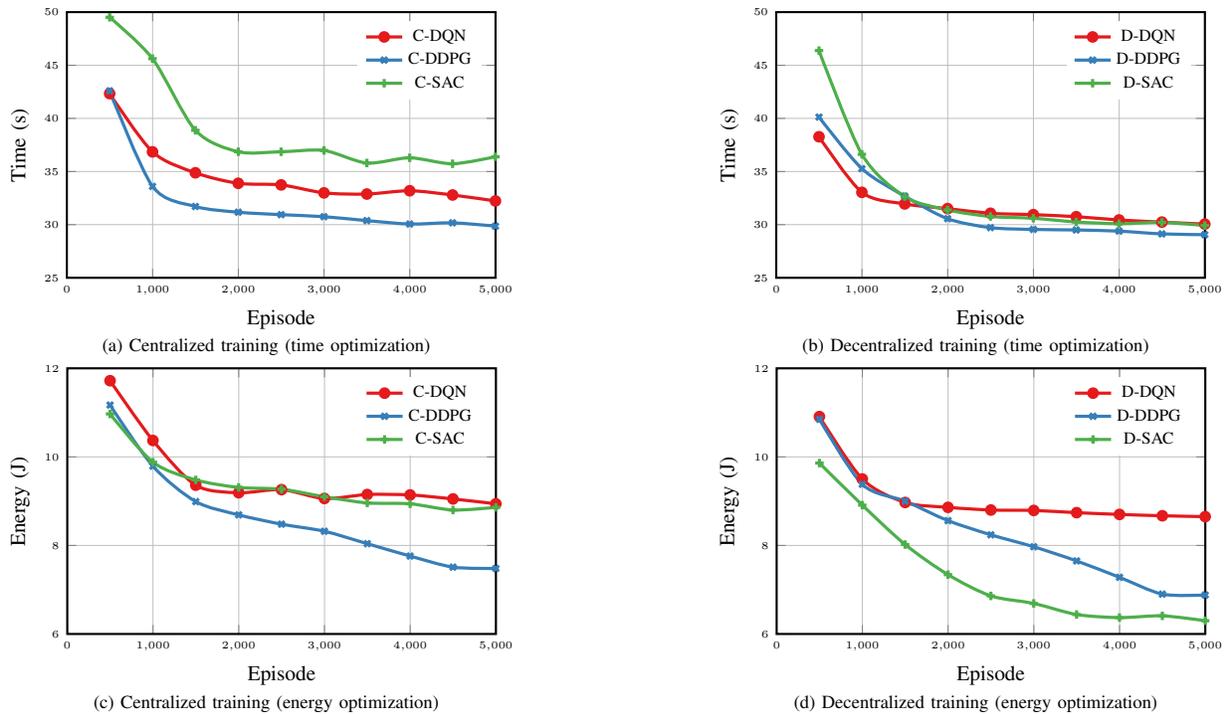
	
	\vspace{-0.3cm}
	\subsection{Simulation Setup and Parameters}
	We consider a vehicular network scenario where a cluster of vehicles travels together on a three-lane road, passing by a roadside base station. The vehicles are initially positioned uniformly between 0-50 m and travel at speeds uniformly distributed between 10-15 m/s. This setup creates a dynamic environment where channel conditions naturally evolve as vehicles move along the road: V2V links are favorable when vehicles are distant from the base station, V2I links become advantageous as vehicles approach the base station, and V2V links regain preference as vehicles move away. This dynamic vehicular scenario provides an ideal testbed for evaluating the adaptability of DRL algorithms, as they must learn to dynamically switch between V2V cooperation and direct V2I transmission in response to time-varying channel conditions.
	
	Table \ref{tab:simulation_params} summarizes the key simulation parameters used in our experiments. These values are selected based on typical vehicular network configurations and 5G NR specifications for V2X communications \cite{3gpp2021nr}. The channel model parameters reflect urban vehicular environments with moderate path loss and fading. For each vehicle, the generated data divided into 30 chunks of 20 Mb, totaling 600 Mb, representing large-scale sensor data or high-resolution video streams captured.

	We compare our DRL algorithms against several baseline approaches to provide comprehensive performance results: i) \textbf{All-Base}: All vehicles upload their entire data directly to the base station without any V2V offloading ($\delta_{i,t} = 0$), using maximum V2I power. This represents the traditional cellular upload approach without cooperation. ii) \textbf{All-Leader}: All vehicles offload their whole data to the leader vehicle for deduplication ($\delta_{i,t} = 1$), using maximum V2V power. This approach maximizes deduplication benefits. iii) \textbf{Balanced}: A fixed strategy where each vehicle splits its data equally between V2V and V2I links ($\delta_{i,t} = 0.5$ for all vehicles), using maximum power on both links. 
	
	\subsection{Training Performance}
	\label{subsec:training_performance}
	We first examine the training convergence of DRL algorithms under both centralized and decentralized frameworks. Fig. \ref{fig:training_curves} represents the evolution of time and energy costs during training for the respective optimization objectives. The training curves demonstrate the learning progress over 5000 episodes, where each point represents a moving average to smooth out the inherent variability. We evaluate DQN, DDPG, and SAC in both centralized (C-DQN, C-DDPG, C-SAC) and decentralized (D-DQN, D-DDPG, D-SAC) configurations.
	
	The training curves reveal distinct patterns across optimization objectives and frameworks. For time optimization, centralized algorithms demonstrate steady convergence with varying characteristics, DQN maintains stable trajectories, DDPG shows moderate variance, and SAC exhibits higher fluctuations due to its stochastic policy sampling. The transition to decentralized implementation yields marginally better performance, with notable stability improvements for SAC while DQN and DDPG show similar convergence patterns. Energy optimization presents noticeably different dynamics, centralized algorithms exhibit slower convergence and higher variance. In contrast, the decentralized framework significantly enhances learning experience, with all algorithms achieving faster convergence and superior final performance. This pronounced difference in energy optimization performance highlights how using local decisions enables agents to adapt to specific channel conditions better.
	
	\begin{table*}[t]
	\caption{Performance evaluation of DRL algorithms for time and energy optimization objectives}
	\centering
	\footnotesize
	\renewcommand{\arraystretch}{1.2}
	\resizebox{\textwidth}{!}{%
	\begin{tabular}{l|>{\columncolor{gray!20}}c|c|c|c|c|c||c|>{\columncolor{gray!20}}c|c|c|c|c}
	\hline
 	& \multicolumn{6}{c||}{\textbf{Time Optimization}} & \multicolumn{6}{c}{\textbf{Energy Optimization}} \\
	\cline{2-13}
	\textbf{Algorithm} & \textbf{Time (s)} & \textbf{Energy (J)} & \boldmath{$\delta_{i,t}$} & \boldmath{$p_i^{\mathrm{v2v}}[t]$} & \boldmath{$p_i^{\mathrm{v2i}}[t]$} & \textbf{Improv.} & \textbf{Time (s)} & \textbf{Energy (J)} & \boldmath{$\delta_{i,t}$} & \boldmath{$p_i^{\mathrm{v2v}}[t]$} & \boldmath{$p_i^{\mathrm{v2i}}[t]$} & \textbf{Improv.} \\
	\hline
	\hline
	All-Leader & 43.55$\pm$5.91 & 14.24$\pm$2.12 & 1.00$\pm$0.00 & 1.00$\pm$0.00 & 0.00$\pm$0.00 & -- & 43.55$\pm$5.91 & 14.24$\pm$2.12 & 1.00$\pm$0.00 & 1.00$\pm$0.00 & 0.00$\pm$0.00 & -- \\
	All-Base & 45.21$\pm$7.52 & 9.04$\pm$1.50 & 0.00$\pm$0.00 & 0.00$\pm$0.00 & 1.00$\pm$0.00 & -- & 45.21$\pm$7.52 & 9.04$\pm$1.50 & 0.00$\pm$0.00 & 0.00$\pm$0.00 & 1.00$\pm$0.00 & 0.0\% \\
	Balanced & 35.86$\pm$5.28 & 11.61$\pm$1.46 & 0.50$\pm$0.00 & 1.00$\pm$0.00 & 1.00$\pm$0.00 & 0.0\% & 35.86$\pm$5.28 & 11.61$\pm$1.46 & 0.50$\pm$0.00 & 1.00$\pm$0.00 & 1.00$\pm$0.00 & -- \\
	\hline
	C-DQN & 31.30$\pm$2.99 & 10.64$\pm$1.25 & 0.48$\pm$0.28 & 0.81$\pm$0.29 & 0.85$\pm$0.23 & 12.7\% & 43.48$\pm$7.03 & 8.86$\pm$1.15 & 0.28$\pm$0.19 & 0.51$\pm$0.20 & 0.70$\pm$0.16 & 2.0\% \\
	C-DDPG & 29.56$\pm$2.94 & 10.39$\pm$0.94 & 0.43$\pm$0.35 & 0.84$\pm$0.25 & 0.97$\pm$0.10 & 17.6\% & 55.16$\pm$8.02 & 7.45$\pm$1.15 & 0.19$\pm$0.29 & 0.26$\pm$0.29 & 0.50$\pm$0.31 & 17.6\% \\
	C-SAC & 31.59$\pm$3.48 & 9.82$\pm$1.13 & 0.48$\pm$0.33 & 0.70$\pm$0.23 & 0.77$\pm$0.20 & 11.9\% & 52.30$\pm$7.75 & 7.69$\pm$1.07 & 0.21$\pm$0.23 & 0.32$\pm$0.26 & 0.52$\pm$0.26 & 14.9\% \\
	\hline
	D-DQN & 29.57$\pm$3.09 & 10.52$\pm$1.37 & 0.48$\pm$0.28 & 0.86$\pm$0.24 & 0.88$\pm$0.20 & 17.5\% & 40.97$\pm$4.29 & 8.64$\pm$1.26 & 0.26$\pm$0.23 & 0.40$\pm$0.28 & 0.74$\pm$0.21 & 4.4\% \\
	D-DDPG & 29.01$\pm$2.56 & 9.47$\pm$1.14 & 0.34$\pm$0.31 & 0.71$\pm$0.35 & 1.00$\pm$0.00 & 19.1\% & 60.55$\pm$13.50 & 7.08$\pm$1.47 & 0.12$\pm$0.28 & 0.28$\pm$0.29 & 0.52$\pm$0.43 & 21.7\% \\
	 \textbf{D-SAC} & \textbf{28.81$\pm$3.27} & 10.10$\pm$1.95 & 0.47$\pm$0.33 & 0.61$\pm$0.19 & 0.85$\pm$0.13 & \textbf{19.7\%} & 70.23$\pm$10.13 & \textbf{6.04$\pm$1.68} & 0.17$\pm$0.28 & 0.31$\pm$0.17 & 0.29$\pm$0.24 & \textbf{33.2\%} \\
	\hline
	\end{tabular}%
	}
	\label{tab:evaluation_results}
	\vspace{-0.4cm}
	\end{table*}

	\subsection{Evaluation Results}
	We evaluate the trained DRL algorithms on 100 test episodes with different vehicular configurations, e.g., different vehicle speeds, positions, and channel conditions, to assess performance under varying network dynamics. Table \ref{tab:evaluation_results} presents the comprehensive performance metrics for both time and energy optimization objectives. For each episode, we compute the total time cost and energy consumption across all time slots, then report the mean and standard deviation across all test episodes. The learned action parameters, offloading fraction $\delta_{i,t}$, V2V transmission power $p_i^{v2v}[t]$, and V2I transmission power $p_i^{v2i}[t]$ are averaged across all follower vehicles and time slots within each episode, then averaged across episodes. We note that these power allocations are represented with normalized values. The improvement percentages are calculated relative to the best-performing baseline algorithm for both time and energy optimization.

	For time optimization, the results reveal a clear and consistent pattern where decentralized approaches significantly outperform their centralized counterparts across all algorithms. D-SAC achieves the best overall performance with 19.7\% improvement over the best baseline (Balanced), while all decentralized variants consistently outperform centralized ones. This consistent superiority of decentralized approaches demonstrates that distributed decision-making provides significant advantages for time optimization. The performance gains stem from the ability of vehicles to make independent decisions based on their local channel observations rather than relying on global coordination. Notably, the learned policies exhibit high power utilization across both V2V and V2I links, indicating that minimizing completion time requires aggressive use of available communication resources. These improvements demonstrate the effectiveness of DRL in learning complex coordination strategies that fixed policies cannot capture in dynamic vehicular offloading scenarios.

	In contrast, energy optimization demonstrates substantially larger performance gains, with improvements of up to 33.2\% over the best baseline algorithm (All-Base). The larger improvements in energy optimization demonstrate that adaptive power control offers substantial benefits against baseline strategies, as DRL algorithms learn to dynamically adjust power levels based on real-time environment conditions. The learned policies reveal a fundamental shift in communication strategy, particularly D-SAC, maintain low power levels across both V2V and V2I links, demonstrating balanced power allocation. This strategy is coupled with significantly reduced V2V offloading fractions, as low power levels make offloading to the leader vehicle less advantageous since deduplication do not compensate for the transmission costs.

\begin{figure}[h]
		\centering
		\begin{tikzpicture}[scale=0.99]
			\begin{axis}[
				width=9cm,
				height=6.5cm,
				xlabel={\small Time Slot},
				ylabel={\small Offloading Fraction ($\delta_{i,t}$)},
				xmin=0, xmax=29,
				ymin=-0.05, ymax=1.05,
				xtick={0,5,10,15,20,25,29},
				ytick={0,0.2,0.4,0.6,0.8,1.0},
				grid=both,
				grid style={line width=.1pt, draw=gray!20},
				major grid style={line width=.2pt,draw=gray!50},
				legend style={
					at={(0.5,-0.2)},
					anchor=north,
					legend columns=3,
					font=\footnotesize,
					fill=white,
					fill opacity=0.9,
					draw=black!50,
					rounded corners=2pt,
					/tikz/every even column/.append style={column sep=0.5cm}
				},
				line width=1pt,
				clip=false,
				axis x line*=bottom,
				axis y line*=left,
				every outer x axis line/.append style={-},
				every outer y axis line/.append style={-},
			]
			\draw[gray!50!black, dashed, line width=1.5pt] (axis cs:13,0) -- (axis cs:13,1);
			\node[gray!60!black, font=\footnotesize, fill=white, fill opacity=0.8] at (axis cs:13,0.85) {Closest to BS};
			
			\addplot[color=purple!70!black, line width=1.2pt, dotted, mark=none, opacity=0.7] coordinates {
				(0,1.0) (1,1.0) (2,1.0) (3,1.0) (4,1.0) (5,1.0) (6,1.0) (7,1.0) (8,1.0) (9,1.0) (10,1.0) (11,1.0) (12,1.0) (13,1.0) (14,1.0) (15,1.0) (16,1.0) (17,1.0) (18,1.0) (19,1.0) (20,1.0) (21,1.0) (22,1.0) (23,1.0) (24,1.0) (25,1.0) (26,1.0) (27,1.0) (28,1.0) (29,1.0)
			};
			\addplot[color=brown!70!black, line width=1.2pt, dotted, mark=none, opacity=0.7] coordinates {
				(0,0.0) (1,0.0) (2,0.0) (3,0.0) (4,0.0) (5,0.0) (6,0.0) (7,0.0) (8,0.0) (9,0.0) (10,0.0) (11,0.0) (12,0.0) (13,0.0) (14,0.0) (15,0.0) (16,0.0) (17,0.0) (18,0.0) (19,0.0) (20,0.0) (21,0.0) (22,0.0) (23,0.0) (24,0.0) (25,0.0) (26,0.0) (27,0.0) (28,0.0) (29,0.0)
			};
			\addplot[color=gray!70!black, line width=1.2pt, dotted, mark=none, opacity=0.7] coordinates {
				(0,0.5) (1,0.5) (2,0.5) (3,0.5) (4,0.5) (5,0.5) (6,0.5) (7,0.5) (8,0.5) (9,0.5) (10,0.5) (11,0.5) (12,0.5) (13,0.5) (14,0.5) (15,0.5) (16,0.5) (17,0.5) (18,0.5) (19,0.5) (20,0.5) (21,0.5) (22,0.5) (23,0.5) (24,0.5) (25,0.5) (26,0.5) (27,0.5) (28,0.5) (29,0.5)
			};
			
			\addplot[color=figcolor1, line width=1.2pt, solid, mark=*, mark size=1.2pt, mark indices={1,5,10,15,20,25,29}, opacity=0.85] coordinates {
				(0,0.7085) (1,0.757) (2,0.757) (3,0.7225) (4,0.6395) (5,0.4898) (6,0.3868) (7,0.337) (8,0.2643) (9,0.2523) (10,0.2305) (11,0.2350) (12,0.2420) (13,0.2500) (14,0.2580) (15,0.2680) (16,0.281) (17,0.3175) (18,0.3478) (19,0.388) (20,0.4138) (21,0.4703) (22,0.5343) (23,0.587) (24,0.6203) (25,0.653) (26,0.6748) (27,0.6965) (28,0.7013) (29,0.698)
			};
			\addplot[color=figcolor2, line width=1.2pt, solid, mark=square*, mark size=1.2pt, mark indices={1,5,10,15,20,25,29}, opacity=0.85] coordinates {
				(0,0.8074) (1,0.8068) (2,0.7746) (3,0.7141) (4,0.6250) (5,0.5105) (6,0.3767) (7,0.2565) (8,0.1577) (9,0.0925) (10,0.0549) (11,0.0423) (12,0.0341) (13,0.0227) (14,0.0229) (15,0.0382) (16,0.0714) (17,0.1232) (18,0.2097) (19,0.3171) (20,0.4380) (21,0.5429) (22,0.6237) (23,0.6819) (24,0.7265) (25,0.7590) (26,0.7821) (27,0.7943) (28,0.7972) (29,0.7927)
			};
			\addplot[color=figcolor3, line width=1.2pt, solid, mark=triangle*, mark size=1.5pt, mark indices={1,5,10,15,20,25,29}, opacity=0.85] coordinates {
				(0,0.784) (1,0.751) (2,0.718) (3,0.709) (4,0.675) (5,0.589) (6,0.469) (7,0.368) (8,0.300) (9,0.236) (10,0.196) (11,0.175) (12,0.162) (13,0.155) (14,0.155) (15,0.179) (16,0.214) (17,0.242) (18,0.289) (19,0.341) (20,0.397) (21,0.449) (22,0.517) (23,0.596) (24,0.681) (25,0.759) (26,0.815) (27,0.843) (28,0.853) (29,0.854)
			};
			\addplot[color=figcolor1!80!black, line width=1.2pt, dashed, mark=o, mark size=1.2pt, mark indices={1,5,10,15,20,25,29}, opacity=0.85] coordinates {
				(0,0.776) (1,0.832) (2,0.794) (3,0.711) (4,0.590) (5,0.507) (6,0.456) (7,0.406) (8,0.361) (9,0.324) (10,0.295) (11,0.236) (12,0.196) (13,0.179) (14,0.178) (15,0.219) (16,0.255) (17,0.301) (18,0.339) (19,0.385) (20,0.436) (21,0.467) (22,0.511) (23,0.563) (24,0.605) (25,0.641) (26,0.674) (27,0.697) (28,0.710) (29,0.712)
			};
			\addplot[color=figcolor2!80!black, line width=1.2pt, dashed, mark=square, mark size=1.2pt, mark indices={1,5,10,15,20,25,29}, opacity=0.85] coordinates {
				(0,0.922) (1,0.823) (2,0.690) (3,0.550) (4,0.394) (5,0.247) (6,0.132) (7,0.064) (8,0.033) (9,0.025) (10,0.023) (11,0.022) (12,0.023) (13,0.025) (14,0.035) (15,0.070) (16,0.135) (17,0.195) (18,0.252) (19,0.302) (20,0.348) (21,0.393) (22,0.438) (23,0.487) (24,0.538) (25,0.583) (26,0.614) (27,0.623) (28,0.604) (29,0.566)
			};
			\addplot[color=figcolor3!80!black, line width=1.2pt, dashed, mark=triangle, mark size=1.5pt, mark indices={1,5,10,15,20,25,29}, opacity=0.85] coordinates {
				(0,0.961) (1,0.888) (2,0.795) (3,0.721) (4,0.638) (5,0.540) (6,0.444) (7,0.342) (8,0.242) (9,0.168) (10,0.122) (11,0.097) (12,0.088) (13,0.090) (14,0.097) (15,0.121) (16,0.167) (17,0.227) (18,0.294) (19,0.359) (20,0.422) (21,0.487) (22,0.552) (23,0.615) (24,0.673) (25,0.724) (26,0.764) (27,0.789) (28,0.794) (29,0.778)
			};
			
			\draw[->, gray!60!black, line width=1.2pt] (axis cs:8,0.95) -- (axis cs:12,0.95);
			\node[gray!60!black, font=\footnotesize] at (axis cs:7,0.92) {Approaching};
			\draw[->, gray!60!black, line width=1.2pt] (axis cs:14,0.95) -- (axis cs:18,0.95);
			\node[gray!60!black, font=\footnotesize] at (axis cs:21,0.92) {Departing};
			
			\draw[gray!60!black, line width=0.8pt] (axis cs:0,1.05) -- (axis cs:29,1.05);
			
			\draw[gray!60!black, line width=0.8pt] (axis cs:0,1.05) -- (axis cs:0,1.08);
			\draw[gray!60!black, line width=0.8pt] (axis cs:5,1.05) -- (axis cs:5,1.08);
			\draw[gray!60!black, line width=0.8pt] (axis cs:10,1.05) -- (axis cs:10,1.08);
			\draw[gray!60!black, line width=0.8pt] (axis cs:15,1.05) -- (axis cs:15,1.08);
			\draw[gray!60!black, line width=0.8pt] (axis cs:20,1.05) -- (axis cs:20,1.08);
			\draw[gray!60!black, line width=0.8pt] (axis cs:25,1.05) -- (axis cs:25,1.08);
			\draw[gray!60!black, line width=0.8pt] (axis cs:29,1.05) -- (axis cs:29,1.08);
			
			\node[gray!60!black, font=\footnotesize] at (axis cs:0,1.12) {162m};
			\node[gray!60!black, font=\footnotesize] at (axis cs:5,1.12) {100m};
			\node[gray!60!black, font=\footnotesize] at (axis cs:10,1.12) {38m};
			\node[gray!60!black, font=\footnotesize] at (axis cs:15,1.12) {31m};
			\node[gray!60!black, font=\footnotesize] at (axis cs:20,1.12) {88m};
			\node[gray!60!black, font=\footnotesize] at (axis cs:25,1.12) {150m};
			\node[gray!60!black, font=\footnotesize] at (axis cs:29,1.12) {200m};
			
			\node[gray!60!black, font=\footnotesize] at (axis cs:14.5,1.22) {Distance to Base Station};
			\legend{All-Leader, All-Base, Balanced, C-DQN, C-DDPG, C-SAC, D-DQN, D-DDPG, D-SAC}
			\end{axis}
		\end{tikzpicture}
		\caption{Evolution of offloading fraction across time slots for time optimization. The top axis shows the average distance of vehicles from the base station, with arrows indicating approaching and departing phases. The vertical dashed line marks the point of closest approach to the base station.}
		\label{fig:v2v_fraction_evolution}
	\end{figure}
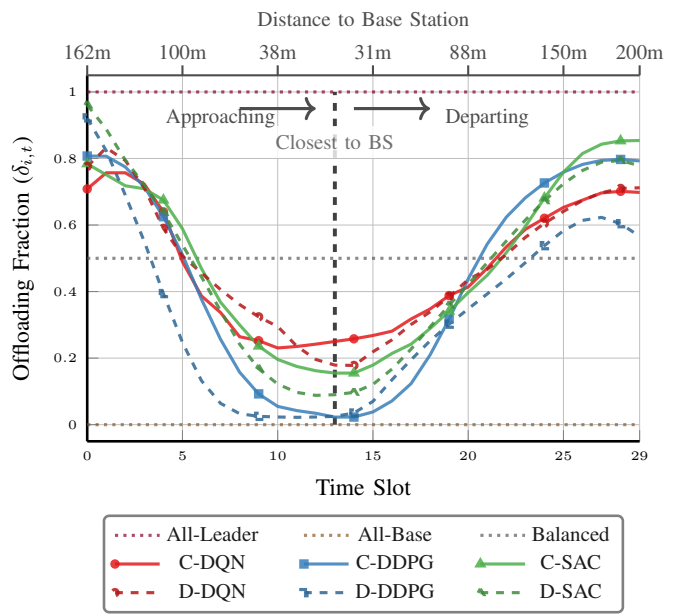

	In order to understand the performance of the algorithms, we examined how they adapt to changing environment conditions throughout an episode. Fig. \ref{fig:v2v_fraction_evolution} illustrates the dynamic adaptation of offloading fraction across time slots. As vehicles approach the base station during the early time slots, DRL algorithms progressively reduce their V2V offloading fraction to the leader vehicle, increasingly favoring direct V2I transmission to exploit the improved channel conditions. After passing the base station, the algorithms demonstrate intelligent adaptation by gradually increasing their V2V offloading fraction again, recognizing that V2V communication becomes more advantageous than V2I as the distance to the base station increases. 

	The U-shaped pattern in offloading behavior reflects the ability of the algorithms to learn the fundamental trade-off between communication modes. The adaptation patterns vary significantly across algorithms, with DDPG variants showing the most aggressive strategy adjustments, nearly eliminating V2V offloading when closest to the base station. SAC variants demonstrate substantial but more controlled adaptation, while DQN maintains the most conservative approach. The baseline algorithms show no adaptation to the changing environment, maintaining their fixed offloading strategies, which explains their inferior performance. \vspace{-0.3cm}

	\begin{table*}[t]
	\caption{Scalability analysis of DRL algorithms across different vehicular cluster sizes for both time and energy optimization objectives.}
	\centering
	\footnotesize
	\renewcommand{\arraystretch}{1.2}
	\resizebox{\textwidth}{!}{%
	\begin{tabular}{l|ccccc||ccccc}
	\hline
	& \multicolumn{5}{c||}{\textbf{Time Optimization (s)}} & \multicolumn{5}{c}{\textbf{Energy Optimization (J)}} \\
	\cline{2-11}
	\textbf{Algorithm} & \textbf{N=3} & \textbf{N=4} & \textbf{N=5} & \textbf{N=6} & \textbf{N=7} & \textbf{N=3} & \textbf{N=4} & \textbf{N=5} & \textbf{N=6} & \textbf{N=7} \\
	\hline
	\hline
	All-Leader & 22.02$\pm$3.40 & 32.96$\pm$4.62 & 43.55$\pm$5.91 & 53.90$\pm$5.92 & 64.45$\pm$7.03 & 7.14$\pm$1.12 & 10.78$\pm$1.75 & 14.24$\pm$2.12 & 17.50$\pm$2.72 & 20.98$\pm$3.22 \\
	All-Base & 22.83$\pm$5.43 & 33.30$\pm$6.50 & 45.21$\pm$7.52 & 56.12$\pm$9.01 & 68.75$\pm$9.17 & 4.57$\pm$1.09 & 6.66$\pm$1.30 & 9.04$\pm$1.50 & 11.22$\pm$1.80 & 13.75$\pm$1.83 \\
	Balanced & 17.70$\pm$2.66 & 25.87$\pm$3.44 & 35.86$\pm$5.28 & 43.72$\pm$4.19 & 52.66$\pm$4.99 & 5.89$\pm$0.75 & 8.65$\pm$1.06 & 11.61$\pm$1.46 & 14.56$\pm$1.46 & 17.50$\pm$1.85 \\
	\hline
	C-DQN & 14.93$\pm$2.00 & 23.16$\pm$2.38 & 31.30$\pm$3.00 & 41.66$\pm$4.36 & 50.59$\pm$4.88 & 4.45$\pm$1.05 & 6.61$\pm$0.94 & 8.86$\pm$1.15 & 11.26$\pm$1.75 & 13.60$\pm$1.58 \\
	C-DDPG & \textbf{13.66$\pm$1.38} & 21.43$\pm$2.17 & 29.56$\pm$2.94 & 39.27$\pm$5.15 & 49.01$\pm$5.62 & 3.23$\pm$0.68 & 5.64$\pm$1.11 & 7.45$\pm$1.15 & 10.97$\pm$1.93 & 12.48$\pm$1.73 \\
	C-SAC & 14.34$\pm$1.20 & 22.30$\pm$1.84 & 31.59$\pm$3.48 & 41.35$\pm$4.45 & 49.26$\pm$5.05 & 3.39$\pm$0.60 & 5.28$\pm$0.91 & 7.69$\pm$1.07 & 10.44$\pm$1.24 & 12.37$\pm$1.34 \\
	\hline
	D-DQN & 14.30$\pm$1.36 & 21.97$\pm$1.69 & 29.57$\pm$3.09 & 36.99$\pm$3.09 & 45.94$\pm$4.12 & 4.19$\pm$0.66 & 6.41$\pm$0.82 & 8.64$\pm$1.26 & 10.79$\pm$1.15 & 13.27$\pm$1.43 \\
	D-DDPG & 14.03$\pm$1.37 & \textbf{20.66$\pm$1.71} & 29.01$\pm$2.56 & 36.84$\pm$3.76 & 46.33$\pm$4.42 & 3.10$\pm$1.01 & 5.20$\pm$0.98 & 7.08$\pm$1.47 & 11.74$\pm$3.10 & 13.81$\pm$1.80 \\
	D-SAC & 14.08$\pm$1.34 & 21.86$\pm$4.83 & \textbf{28.81$\pm$3.27} & \textbf{35.78$\pm$2.57} & \textbf{44.34$\pm$3.52} & \textbf{2.92$\pm$0.58} & \textbf{4.71$\pm$2.71} & \textbf{6.04$\pm$1.68} & \textbf{8.00$\pm$0.97} & \textbf{9.47$\pm$1.16} \\
	\hline
	\end{tabular}%
	}
	\label{tab:scalability_results}
	\end{table*}
	
	\begin{table*}[t]
	\caption{Performance evaluation of DRL algorithms under varying redundancy ratios for both time and energy optimization objectives.}
	\centering
	\footnotesize
	\renewcommand{\arraystretch}{1.2}
	\resizebox{\textwidth}{!}{%
	\begin{tabular}{l|ccccc||ccccc}
	\hline
	& \multicolumn{5}{c||}{\textbf{Time Optimization (s)}} & \multicolumn{5}{c}{\textbf{Energy Optimization (J)}} \\
	\cline{2-11}
	\textbf{Algorithm} & \textbf{$\boldsymbol{\beta}_{i,t}$=0.3} & \textbf{$\boldsymbol{\beta}_{i,t}$=0.4} & \textbf{$\boldsymbol{\beta}_{i,t}$=0.5} & \textbf{$\boldsymbol{\beta}_{i,t}$=0.6} & \textbf{$\boldsymbol{\beta}_{i,t}$=0.7} & \textbf{$\boldsymbol{\beta}_{i,t}$=0.3} & \textbf{$\boldsymbol{\beta}_{i,t}$=0.4} & \textbf{$\boldsymbol{\beta}_{i,t}$=0.5} & \textbf{$\boldsymbol{\beta}_{i,t}$=0.6} & \textbf{$\boldsymbol{\beta}_{i,t}$=0.7} \\
	\hline
	\hline
	All-Leader & 48.38$\pm$6.39 & 45.97$\pm$6.14 & 43.55$\pm$5.91 & 41.13$\pm$5.69 & 38.72$\pm$5.50 & 18.10$\pm$2.71 & 16.17$\pm$2.41 & 14.24$\pm$2.12 & 12.30$\pm$1.84 & 10.37$\pm$1.57 \\
	All-Base & 45.21$\pm$7.52 & 45.21$\pm$7.52 & 45.21$\pm$7.52 & 45.21$\pm$7.52 & 45.21$\pm$7.52 & 9.04$\pm$1.50 & 9.04$\pm$1.50 & 9.04$\pm$1.50 & 9.04$\pm$1.50 & 9.04$\pm$1.50 \\
	Balanced & 37.25$\pm$5.43 & 36.05$\pm$5.35 & 35.86$\pm$5.28 & 33.66$\pm$5.22 & 32.47$\pm$5.17 & 13.53$\pm$1.70 & 12.57$\pm$1.57 & 11.61$\pm$1.46 & 10.66$\pm$1.36 & 9.70$\pm$1.27 \\
	\hline
	C-DQN & 35.96$\pm$4.11 & 33.33$\pm$3.24 & 31.30$\pm$3.00 & 29.34$\pm$2.62 & 27.82$\pm$2.63 & 9.06$\pm$1.32 & 9.07$\pm$1.23 & 8.86$\pm$1.15 & 8.33$\pm$1.02 & 7.53$\pm$0.95 \\
	C-DDPG & 34.42$\pm$3.89 & 31.17$\pm$3.15 & 29.56$\pm$2.94 & 27.74$\pm$2.63 & 26.11$\pm$2.69 & 8.28$\pm$1.27 & 8.20$\pm$1.21 & 7.45$\pm$1.15 & 6.85$\pm$0.88 & 6.07$\pm$0.83 \\
	C-SAC & 36.03$\pm$4.09 & 33.89$\pm$3.71 & 31.59$\pm$3.48 & 29.26$\pm$2.63 & 27.45$\pm$2.63 & 8.44$\pm$1.25 & 7.93$\pm$1.20 & 7.69$\pm$1.07 & 6.85$\pm$0.91 & 5.95$\pm$0.87 \\
	\hline
	D-DQN & 33.57$\pm$3.25 & 31.79$\pm$3.57 & 29.57$\pm$3.09 & 27.42$\pm$2.38 & 25.97$\pm$2.36 & 9.25$\pm$1.41 & 9.15$\pm$1.28 & 8.64$\pm$1.26 & 7.99$\pm$1.17 & 7.15$\pm$1.00 \\
	D-DDPG & \textbf{31.81$\pm$3.30} & \textbf{29.62$\pm$2.47} & 29.01$\pm$2.56 & 29.28$\pm$2.75 & \textbf{24.85$\pm$1.87} & 8.02$\pm$1.71 & 9.14$\pm$1.80 & 7.08$\pm$1.47 & 5.78$\pm$1.02 & 5.61$\pm$0.68 \\
	D-SAC & 32.85$\pm$4.14 & 31.01$\pm$3.93 & \textbf{28.81$\pm$3.27} & \textbf{27.08$\pm$2.82} & 25.40$\pm$2.51 & \textbf{6.97$\pm$1.49} & \textbf{6.96$\pm$1.55} & \textbf{6.04$\pm$1.68} & \textbf{5.70$\pm$1.48} & \textbf{4.96$\pm$1.15} \\
	\hline
	\end{tabular}%
	}
	\label{tab:redundancy_results}
	\vspace{-0.4cm}
	\end{table*}

	We evaluate the scalability of our approach beyond the baseline 5 vehicle configuration by conducting comprehensive experiments across cluster sizes ranging from 3 to 7 vehicles. Table \ref{tab:scalability_results} presents detailed performance metrics for both time and energy optimization objectives, with each configuration evaluated over 100 episodes to ensure reliable comparisons.
	
	The scalability analysis reveals distinct patterns in algorithm performance with different cluster sizes. For time optimization, the results show an interesting progression where DDPG variants demonstrate superior performance for smaller cluster sizes, as the deterministic policy gradients provide more stable learning in simpler coordination scenarios. However, as cluster size increases, D-SAC achieves marginally better performance, with D-DDPG maintaining competitive results, suggesting that both approaches remain viable for larger vehicular networks. Comparing centralized and decentralized approaches, decentralized variants consistently outperform their centralized counterparts across all cluster sizes. For energy optimization, D-SAC demonstrates remarkable consistency, achieving the best performance across all cluster sizes. This consistent advantage highlights the effectiveness of entropy-regularized learning. The centralized and decentralized comparison shows similar trends, where decentralized algorithms maintain their superiority, particularly for larger vehicular networks. The performance gaps between DRL algorithms and baselines widen significantly with larger clusters, demonstrating that adaptive decisions become more valuable as system complexity grows. 

	We further investigate the impact of redundancy ratio $\beta_{i,t}$ on algorithm performances by conducting experiments with varying redundancy levels from 0.3 to 0.7. The redundancy ratio quantifies the proportion of duplicate data content among chunks of vehicles, reflecting the overlap in data captured by vehicles in close proximity. Table \ref{tab:redundancy_results} presents the performance metrics across different redundancy ratios.

	The redundancy ratio analysis reveals how algorithms perform as data duplication levels vary. For time optimization, all algorithms show improved performance as redundancy ratio increases, since higher redundancy means more duplicate data can be eliminated through deduplication at the leader, reducing the total amount of data that needs to be uploaded to the base station. DRL algorithms demonstrate superior performance to varying redundancy levels compared to baselines, with D-DDPG and D-SAC achieving the best performance across different redundancy ratios. As expected, the All-Base baseline shows no variation with redundancy ratio changes, as it exclusively uses V2I communication.

	For energy optimization, the performance improvements with increasing redundancy ratio are even more significant. Higher redundancy ratios make V2V offloading more attractive as the energy cost of transmitting to the leader is offset by the reduced data volume that ultimately needs to reach the base station. The DRL algorithms achieve substantial energy savings especially at higher redundancy ratios, with D-SAC consistently achieving the best performance across all redundancy levels, while D-DDPG provides very close performance values. These algorithms achieve the lowest energy consumption and effectively learn to exploit the trade-off between V2V transmission costs and the benefits of deduplication.

	\section{Conclusions}
	\label{sec:conclusions}
	In this paper, we have proposed the utilization of deep reinforcement learning for data offloading in vehicular networks. Our focus has been on optimizing data distribution between V2V and V2I links while leveraging deduplication capabilities at the leader vehicle to eliminate the redundant content from overlapping data. We established a mathematical framework and formulated two separate optimization problems, one for minimizing total time and one for minimizing total energy consumption. Subsequently, we leveraged the capabilities of deep reinforcement learning algorithms (DQN, DDPG, and SAC) in both centralized and decentralized settings to address these optimization problems under dynamic vehicular network conditions. Through extensive simulations, we have demonstrated that decentralized DRL algorithms, provide superior performance compared to both centralized approaches and baseline strategies across various network conditions and redundancy levels. Moreover, our results highlight the substantial benefits of V2V data offloading with deduplication, as it significantly reduces both time and energy consumption while alleviating congestion on cellular networks.

	\section*{Acknowledgment}
	We would like to thank Mohsen Bahrami, Hamed Asadi, Tengchan Zeng, Yun-Ho Lee, and Basavaraj Tonshal for very helpful discussions. 
	
\bibliographystyle{IEEEtran}
\bibliography{ref}
	
\end{document}